\documentclass[trackchanges,astrosymb,twocolumn]{aastex701}
\setcitestyle{sort}

\hypersetup{linkcolor=blue,citecolor=blue,filecolor=cyan,urlcolor=blue}

\usepackage[dvipsnames]{xcolor}
\usepackage{tabularx}

\newcommand{\nfrbsample}{1683\xspace}
\newcommand{\nfrbcat}{4539\xspace}
\newcommand{\nsources}{3641\xspace}

\newcommand{\catii}{Catalog 2\xspace}
\newcommand{\cati}{Catalog 1\xspace}

\newcommand{\dmcosmic}{\text{DM}_{\text{cosmic}}}
\newcommand{\dmunits}{\text{pc~cm}^{-3}}
\newcommand{\dmhost}{\text{DM}_{\text{host}}}

\newcommand{\fitburst}[0]{{\tt fitburst}\xspace}
\newcommand{\dm}{\ensuremath{\text{DM}}\xspace}
\newcommand{\dmeg}{\ensuremath{\text{DM}_{\text{EG}}}\xspace}
\newcommand{\dmmw}{\overline{\text{DM}}_{\text{MW}}}
\usepackage{xspace}
\newcommand{\start}{25 July 2018\xspace}
\newcommand{\finish}{15 September 2023\xspace}

\defcitealias{chimefrbcatalog1}{Catalog~1}
\defcitealias{abb+25}{Catalog~2}

\newcommand{\snr}{\ensuremath{\text{S/N}}\xspace}

\usepackage{amsmath}
\usepackage{amssymb}
\usepackage{physics}
\usepackage{chngcntr}

\usepackage{threeparttable}

\graphicspath{{./}{}}

\begin{document}

% \title{Modeling the Energy and Redshift distribution of Fast Radio Bursts using the Second CHIME/FRB catalog}
% \title{Modeling the FRB Population Using the Second CHIME/FRB Catalog }
\title{The FRB Population Revisited: Rate, Energy, Evolution, and Host-DM Constraints from the Second CHIME/FRB Catalog}

% \correspondingauthor{Naman Jain}
% \email{naman.jain@mail.mcgill.ca}

\author[0009-0009-0938-1595]{Naman Jain}
\email{naman.jain@mail.mcgill.ca}
\affiliation{Trottier Space Institute at McGill University, 3550 rue University, Montr\'eal, QC H3A 2T7, Canada}
\affiliation{Department of Physics, McGill University, 3600 rue University, Montr\'eal, QC H3A 2T8, Canada}

\author[0000-0002-6823-2073]{Kaitlyn Shin}
\email{kaitshin@caltech.edu}
\affiliation{Cahill Center for Astronomy and Astrophysics, MC 249-17 California Institute of Technology, Pasadena CA 91125, USA}

\author[0000-0002-4623-5329]{Mawson Sammons}
\email{mawson.sammons@lam.fr}
\affiliation{Laboratoire d'Astrophysique de Marseille, Aix-Marseille Univ., CNRS, CNES, Marseille, France}

\author[0000-0003-2111-3437]{Kyle McGregor}
\email{kyle.mcgregor@mail.mcgill.ca}
\affiliation{Trottier Space Institute at McGill University, 3550 rue University, Montr\'eal, QC H3A 2T7, Canada}
\affiliation{Department of Physics, McGill University, 3600 rue University, Montr\'eal, QC H3A 2T8, Canada}

\author[0000-0001-9345-0307]{Victoria Kaspi}
\email{victoria.kaspi@mcgill.ca}
\affiliation{Trottier Space Institute at McGill University, 3550 rue University, Montr\'eal, QC H3A 2T7, Canada}
\affiliation{Department of Physics, McGill University, 3600 rue University, Montr\'eal, QC H3A 2T8, Canada}

\author[0000-0002-4279-6946]{Kiyoshi W.~Masui}
\email{kmasui@mit.edu}
\affiliation{MIT Kavli Institute for Astrophysics and Space Research, Massachusetts Institute of Technology, 77 Massachusetts Ave, Cambridge, MA 02139, USA}
\affiliation{Department of Physics, Massachusetts Institute of Technology, 77 Massachusetts Ave, Cambridge, MA 02139, USA}

\author[0000-0002-8376-1563]{Alice P.~Curtin}
\email{alice.curtin@mcgill.ca}
\affiliation{Department of Physics, McGill University, 3600 rue University, Montr\'eal, QC H3A 2T8, Canada}
\affiliation{Trottier Space Institute at McGill University, 3550 rue University, Montr\'eal, QC H3A 2T7, Canada}
\affiliation{Department of Astronomy, University of Virginia, Charlottesville, VA 22904, USA}

\author[0000-0001-7166-6422]{Matt Dobbs}
\email{matt.dobbs@mcgill.ca}
\affiliation{Department of Physics, McGill University, 3600 rue University, Montr\'eal, QC H3A 2T8, Canada}
\affiliation{Trottier Space Institute at McGill University, 3550 rue University, Montr\'eal, QC H3A 2T7, Canada}

\author[0000-0002-3382-9558]{B.~M.~Gaensler}
\email{gaensler@ucsc.edu}
\affiliation{Department of Astronomy \& Astrophysics, University of California, Santa Cruz, 1156 High Street, Santa Cruz, CA 95064, USA}
\affiliation{Dunlap Institute for Astronomy \& Astrophysics, University of Toronto, 50 St.~George Street, Toronto, ON M5S 3H4, Canada}
\affiliation{David A.\ Dunlap Department of Astronomy \& Astrophysics, University of Toronto, 50 St.~George Street, Toronto, ON M5S 3H4, Canada}

\author[0000-0003-2317-1446]{Jason W. T. Hessels}
\email{jason.hessels@mcgill.ca}
\affiliation{Department of Physics, McGill University, 3600 rue University, Montr\'eal, QC H3A 2T8, Canada}
\affiliation{Trottier Space Institute at McGill University, 3550 rue University, Montr\'eal, QC H3A 2T7, Canada}
\affiliation{Anton Pannekoek Institute for Astronomy, University of Amsterdam, Science Park 904, 1098 XH, Amsterdam, The Netherlands}
\affiliation{ASTRON, Netherlands Institute for Radio Astronomy, Oude Hoogeveensedijk 4, 7991 PD Dwingeloo, The Netherlands}

\author[0009-0004-4176-0062]{Afrokk Khan}
\email{afrasiyab.khan@mcgill.ca}
\affiliation{Trottier Space Institute at McGill University, 3550 rue University, Montr\'eal, QC H3A 2T7, Canada}
\affiliation{Department of Physics, McGill University, 3600 rue University, Montr\'eal, QC H3A 2T8, Canada}

\author[0000-0003-2116-3573]{Adam E. Lanman}
\email{alanman@mit.edu}
\affiliation{MIT Kavli Institute for Astrophysics and Space Research, Massachusetts Institute of Technology, 77 Massachusetts Ave, Cambridge, MA 02139, USA}
\affiliation{Department of Physics, Massachusetts Institute of Technology, 77 Massachusetts Ave, Cambridge, MA 02139, USA}

\author[0000-0002-4209-7408]{Calvin Leung}
\email{calvin_leung@berkeley.edu}
\affiliation{Miller Institute for Basic Research, University of California, Berkeley, CA 94720, United States}
\affiliation{Department of Astronomy, University of California, Berkeley, CA 94720, United States}

\author[0000-0003-4584-8841]{Lluis Mas-Ribas}
\email{lmasriba@ucsc.edu}
\affiliation{Department of Astronomy \& Astrophysics, University of California, Santa Cruz, 1156 High Street, Santa Cruz, CA 95064, USA}

\author[0000-0002-2551-7554]{Daniele Michilli}
\email{danielemichilli@gmail.com}
\affiliation{Laboratoire d'Astrophysique de Marseille, Aix-Marseille Univ., CNRS, CNES, Marseille, France}

\author[0000-0002-0772-9326]{Juan Mena-Parra}
\email{juan.menaparra@utoronto.ca}
\affiliation{Dunlap Institute for Astronomy \& Astrophysics, University of Toronto, 50 St.~George Street, Toronto, ON M5S 3H4, Canada}
\affiliation{David A.\ Dunlap Department of Astronomy \& Astrophysics, University of Toronto, 50 St.~George Street, Toronto, ON M5S 3H4, Canada}

\author[0000-0003-0510-0740]{Kenzie Nimmo}
\email{knimmo@northwestern.edu}
\affiliation{Center for Interdisciplinary Exploration and Research in Astrophysics (CIERA), Northwestern University, 1800 Sherman Ave, Evanston, IL 60201, USA}

\author[0000-0002-8897-1973]{Ayush Pandhi}
\email{ayush.pandhi@mcgill.ca}
\affiliation{Trottier Space Institute at McGill University, 3550 rue University, Montr\'eal, QC H3A 2T7, Canada}
\affiliation{Department of Physics, McGill University, 3600 rue University, Montr\'eal, QC H3A 2T8, Canada}

\author[0000-0002-8912-0732]{Aaron B. Pearlman}
\email{aaron.b.pearlman@mit.edu}
\altaffiliation{NASA Hubble Fellow.}
\affiliation{MIT Kavli Institute for Astrophysics and Space Research, Massachusetts Institute of Technology, 77 Massachusetts Ave, Cambridge, MA 02139, USA}
\affiliation{Department of Physics, McGill University, 3600 rue University, Montr\'eal, QC H3A 2T8, Canada}
\affiliation{Trottier Space Institute at McGill University, 3550 rue University, Montr\'eal, QC H3A 2T7, Canada}

\author[0000-0002-4795-697X]{Ziggy Pleunis}
\email{z.pleunis@uva.nl}
\affiliation{Anton Pannekoek Institute for Astronomy, University of Amsterdam, Science Park 904, 1098 XH, Amsterdam, The Netherlands}
\affiliation{ASTRON, Netherlands Institute for Radio Astronomy, Oude Hoogeveensedijk 4, 7991 PD Dwingeloo, The Netherlands}

\author[0000-0002-8823-8835]{Aniket Prasad}
\email{aniket.prasad@lam.fr}
\affiliation{Laboratoire d'Astrophysique de Marseille, Aix-Marseille Univ., CNRS, CNES, Marseille, France}

\author[0000-0002-7738-6875]{J.~Xavier Prochaska}
\affiliation{Department of Astronomy and Astrophysics, University of California, Santa Cruz, 1156 High Street, Santa Cruz, CA 95060, USA}
\email{xavier@ucolick.org}

\author[0000-0002-7374-7119]{Paul Scholz}
\email{pscholz@yorku.ca}
\affiliation{Department of Physics and Astronomy, York University, 4700 Keele Street, Toronto, Ontario, ON M3J 1P3, Canada}

\author[0000-0002-4823-1946]{Vishwangi Shah}
\email{vishwangi.shah@mail.mcgill.ca}
\affiliation{Trottier Space Institute at McGill University, 3550 rue University, Montr\'eal, QC H3A 2T7, Canada}
\affiliation{Department of Physics, McGill University, 3600 rue University, Montr\'eal, QC H3A 2T8, Canada}

\author[0000-0003-2631-6217]{Seth R.~Siegel}
\email{Seth.Siegel@skao.int}
\affiliation{SKA Observatory, SKA-Low Science Operations Centre, 26 Dick Perry Avenue, Kensington, WA 6151, Australia}
\affiliation{Perimeter Institute for Theoretical Physics, 31 Caroline Street N, Waterloo, ON N25 2YL, Canada}
\affiliation{Department of Physics, McGill University, 3600 rue University, Montr\'eal, QC H3A 2T8, Canada}
\affiliation{Trottier Space Institute at McGill University, 3550 rue University, Montr\'eal, QC H3A 2T7, Canada}

\author[0000-0002-2088-3125]{Kendrick Smith}
\email{kmsmith@perimeterinstitute.ca}
\affiliation{Perimeter Institute for Theoretical Physics, 31 Caroline Street N, Waterloo, ON N25 2YL, Canada}

\author[0000-0001-9784-8670]{Ingrid Stairs}
\email{stairs@astro.ubc.ca}
\affiliation{Department of Physics and Astronomy, University of British Columbia, 6224 Agricultural Road, Vancouver, BC V6T 1Z1 Canada}

\author[0000-0002-5815-6548]{Habtamu Menberu Tedila}
\email{habtamu.tedila@mcgill.ca}
\affiliation{Department of Physics, McGill University, 3600 rue University, Montr\'eal, QC H3A 2T8, Canada}
\affiliation{Trottier Space Institute at McGill University, 3550 rue University, Montr\'eal, QC H3A 2T7, Canada}

\author[0009-0001-3334-9482]{Michele Woodland}
\email{miwoodla@ucsc.edu}
\affiliation{Department of Astronomy \& Astrophysics, University of California, Santa Cruz, 1156 High Street, Santa Cruz, CA 95064, USA}

% \collaboration{all}{CHIME/FRB collaboration}
 
\begin{abstract}
We present a study that simultaneously fits for key fast radio burst (FRB) population properties using the Second CHIME/FRB Catalog while calibrating for instrumental selection effects.
Our inferred volumetric rate at $z=0$ is {$1.1^{+0.4}_{-0.3}\times 10^{5}~{\rm Gpc}^{-3}~{\rm yr}^{-1}$} above an {isotropic-equivalent} energy of $10^{39}$~erg and below a scattering timescale of 30~ms at 600~MHz. The evolution of the FRB rate relative to the cosmic star formation rate remains statistically unconstrained. 
Assuming a Schechter function for the energy distribution, isotropic emission, and 1-GHz emission bandwidth, we find a characteristic maximum energy of $2.0^{+2.6}_{-1.0} \times 10^{41}$~erg and a differential power-law index of {$-1.41\,^{+0.27}_{-0.21}({\rm stat.})\,\pm\,0.35({\rm sys.})$}.
The host galaxy contribution to the observed dispersion measure (DM) is modeled as a log-normal distribution; assuming a total Galactic contribution of 80~$\dmunits$, we find a median host-galaxy DM contribution of $199^{+34}_{-36}\;\dmunits$ {in the rest-frame}. 
Finally, we find that the average FRB spectrum, described by a spectral index of $\alpha=-1.0^{+0.5}_{-0.6}$, is inconsistent with being flat at the $\sim2\sigma$ level.
We note an excess of high-\snr bursts at low DMs in the observed sample that is not predicted by the best-fit model. Testing alternative models does not alleviate this tension; future population studies may mitigate this limitation by incorporating robust fluence and redshift measurements.
\end{abstract}

\keywords{\uat{Radio transient sources}{2008}}

\section{Introduction} 

Since their discovery by \cite{lbm+07}, fast radio bursts (FRBs) have proved to be a puzzling phenomenon. Clearly ubiquitous in the sky, with all-sky rates of hundreds per day \citep[e.g.,][]{chimefrbcatalog1}, the origins of FRBs remain unknown despite extensive study over nearly 20 years. 
% While we observe $\sim3\%$ of FRB sources to repeat \citep[e.g.,][]{ssh+16a,abb+19b,abb+23,rn4}, the true fraction of repeaters remains unknown, and it is unclear whether repeaters represent an independent class of sources. {\citet{jam23} and \citet{rn4} find that a single population of repeating FRBs following certain repetition-rate distributions can reproduce the observed population of non-repeaters. Nevertheless, the two classes continue to exhibit distinct morphologies \citep[e.g.,][]{csr+19, pgk+21, csp+25}, suggesting that they may stem from distinct physical origins.}
{
Roughly 3\% of the FRB sources observed in \catii repeat. While repeaters are observed to have longer durations and narrower bandwidths compared to apparent non-repeaters, suggesting the two may represent distinct populations \citep[e.g.,][]{csr+19, pgk+21, csp+25}, other works suggest that apparent non-repeaters are consistent with having the same underlying energy function as repeaters \citep{koh+24,och+26,shk+26}.
Additionally, \citet{jam23} and \citet{rn4} find that a single population of repeating FRBs following certain repetition-rate distributions can reproduce the observed population of non-repeaters.
}
%One Galactic magnetar, SGR 1935+2154, was seen to emit an FRB-like event \citep{abb+20,brb+20}, proving that magnetars can emit radio bursts comparable to some observed FRBs. However, it remains unclear what fraction of FRBs may originate from magnetars. Similarly, recently one apparently rotation-powered pulsar has been shown to emit bursts of luminosity that rival those of some FRBs \citep{drt+26}, but again, it remains unclear whether all or even most FRBs have such an origin. Associations with host galaxies have also not fully converged on even the general age of FRBs progenitrors; many have been associated with star-forming galaxies \citep{gfk+23,src+24,bmk+23}, but at least two FRBs are in clearly non-star-forming regions \citep{kmn+22,ssl+24,edf+25}.

Population modeling is an important key to unlocking the origin and nature of FRBs. Through careful study of a large number of FRBs, essential properties of FRBs such as their redshift distribution, luminosity or energy function, and volumetric rate can be determined, providing major clues to their origin. Moreover, in such studies FRB host galaxy properties such as dispersion measure (DM) contribution can simultaneously be determined, which is important for realizing FRBs’ potential as cosmic probes \citep[e.g.,][]{rav18,mpm+20,wgf+25}. To maximize constraining power while minimizing systematic and statistical errors, such analyses are best done on a large population found with a single telescope-observing system having a well-determined instrumental and software selection function, thereby mitigating biases against FRBs {populating certain regions of the parameter phase space}. 

Multiple studies attempted to model the FRB population using the Commensal Real-time ASKAP Fast Transients \citep[CRAFT;][]{mbb+10} and Parkes/Murriyang samples at $\sim$GHz frequencies \citep[e.g.,][]{me18b,niino18,jem+19,lp19,lml+20,zzl+20}. These works provided initial constraints on population properties, including the volumetric rate and luminosity function \citep{lml+20}, source-count distribution \citep{jem+19}, and DM distributions \citep{ams+21}. 
\cite{jpm+22a} modeled the FRB population with a joint redshift-dispersion measure distribution (\textit{z}DM) to demonstrate that the cosmic volumetric rate of FRBs evolves over cosmic time, while simultaneously determining the host galaxy contribution to the DM budget, and cumulative slope and maximum energy for the power-law energy function.
\cite{jpm+22b} showed that at higher dispersion measures, complex selection effects invert the $\text{DM}-z$ relation \citep{mpm+20}, indicating that the highest-DM unlocalized FRBs are unlikely to be the most distant.
Finally, \cite{jgp+22c} showed that FRBs with host redshifts are instrumental in FRB population modeling. 

The First CHIME/FRB catalog \citep[][hereafter \cati]{chimefrbcatalog1}, containing 536 events, presented a significant step forward in the field by increasing the number of known FRBs by an order of magnitude. It was accompanied by a well-determined instrumental selection function determined via injections of $\sim$85,000 synthetic FRBs into the CHIME/FRB detection pipeline \citep{mts+23}. 
Using this dataset, \citet[][hereafter referred to as S23]{smb+23} compared selection-corrected populations, generated using the model described by \cite{jpm+22a}, with the observed data by evaluating a Bayesian Poisson-likelihood fit, measuring robust population parameters describing FRBs observed by CHIME/FRB.

Many subsequent works have attempted to model the FRB population with \cati to further constrain its energy function \citep[e.g.,][]{hgc+22, zlz+24, aje+25, lwd+25}, formation rates and redshift evolution \citep[e.g.,][]{qlw+22, zz2022, cjd+24, llz+24, wv24, cp25, zlz+25}.
However there is no consensus on what fraction of the FRB population tracks star formation \citep{wv24}. In contrast, some fraction of the FRB population may follow a delayed channel relative to star formation \citep{qlw+22, src+24}, {supported by the localization of an FRB to a globular cluster} \citep{bgk+21, kmn+22}, or potentially trace stellar mass density \citep{hgc+22, lls+25, hm26}. 
%In one example, \cite{gv21} performed a multi-survey fit including \cati {along with Parkes-HTRU, ASKAP-Incoh, and WSRT-Apertif} and found an optimal population reproducing DM, \snr, and detection rates across all four surveys, concluding that the population is broadly consistent with a magnetar-origin scenario, suggesting an underlying young stellar population. 
The range of conflicting results likely reflects the diversity of underlying population parameter assumptions, observed population selection, fitting techniques, and handling of selection biases by these various authors. In particular, CHIME/FRB has significant selection biases \citep{mts+23, mhj+26} that must be taken into account.
% For example, \cite{qlw+22} found, for a fixed energy cutoff, that the population showed better evidence for following a delayed channel relative to star formation, than star formation itself. 
% \cite{hgc+22} used the non-parametric V/Vmax method on \cati to derive the FRB energy function and its redshift evolution, finding evidence that FRB rate tracks stellar mass density over instantaneous star-formation rate, favoring progenitors with a delay relative to star formation. 
% In contrast, \cite{wv24} looked at the non-repeating sources in \cati and concluded that the rate does indeed follow cosmic star formation. 

Here, we expand on the analysis set out by S23 for the Second CHIME/FRB catalog \citep[][hereafter \catii]{abb+25}, which contains 4539 events, nearly an order of magnitude more than in \cati. It is accompanied by a new measurement of the CHIME/FRB selection function that extends our constraint on the selection bias against scattering time from 10~ms to 30~ms at 600~MHz \citep{mhj+26}. We summarize the catalog and injections in \S\ref{sec:data}. Our goals here are to leverage this unparalleled data set and improved selection function to better constrain the various key FRB population properties, using the FRB population model presented in \S\ref{sec:pop_model} and the methodology described in \S\ref{sec:method} -- comparing the observed population with selection-corrected simulated populations and fitting with Poisson likelihood evaluated on an S/N--DM grid via Markov Chain Monte Carlo (MCMC). We present our best-fit results, along with a goodness-of-fit evaluation and a sensitivity analysis to the assumed Milky Way DM contribution, in \S\ref{sec:results}, and discuss these results in the context of FRB astrophysics in \S\ref{sec:discussion}. Finally, in \S\ref{sec:future} we consider the outlook for future FRB population analysis.

\section{Data}
\label{sec:data}

The Canadian Hydrogen Intensity Mapping Experiment (CHIME) is located at the Dominion Radio Astrophysical Observatory (DRAO) in Penticton, Canada. It comprises four immovable $20\,\rm{m} \times 100\,\rm{m}$ parabolic cylinders, each oriented N-S and populated with 256 dual-linear-polarization feeds sensitive to the frequency range $400-800\,\rm{MHz}$. It scans the entire northern sky above declination $\delta>-11^\circ$ each day. A detailed description of the CHIME telescope is provided by \cite{abb+22}. Its broad frequency coverage, large instantaneous field of view ($\sim$200 square degrees), and daily scans of the full northern sky make CHIME an excellent FRB search instrument. The CHIME/FRB collaboration performs a real-time FRB search through the CHIME/FRB backend; details of the instrument and search algorithm is provided by \cite{abb+18}. 

The following subsections describe the sample of FRBs selected from the Second CHIME/FRB Catalog\footnote[1]{The data can be accessed at \url{https://www.chime-frb.ca/catalog2}.} and the revised CHIME/FRB injection campaign, following \cite{mhj+26}, used to characterize instrumental selection effects.

\subsection{The Second CHIME/FRB Catalog}
\label{sec:catalog}

The second CHIME/FRB catalog contains \nfrbcat bursts from \nsources unique sources observed between {\start} and {\finish}. CHIME/FRB employs the \fitburst framework \citep{fpb+24}, a least-squares routine, to generate empirical models of the dynamic spectra of all detected FRBs. This fitting routine produces measurements for the arrival time ($t_\text{arr}$), amplitude ($A_i$), intrinsic temporal width ($w$), scattering timescale ($\tau $, referenced to 600~MHz), spectral index\footnote[2]{Spectral index is denoted as $\gamma $ by \catii; we follow the notation of S23.} ($\alpha$), spectral running ($r$), and \dm. For bursts without a significant scattering measurement, we adopt $w$ as an upper-limit value for $\tau$.

For multi-component FRBs for which each sub-burst $i$ has distinct parameters, we retain only the first sub-burst ($i=0$) and its \fitburst{} width is replaced with an equivalent proxy width derived from a linear relationship between boxcar and \fitburst{} widths calibrated on single-component bursts.

{Because Catalog 2 fluence values are strict lower limits due to large localization uncertainty with respect to the telescope beam pattern} \citep{apb+23}, we follow S23 and instead use the \snr from the real-time detection pipeline \citep[\texttt{bonsai};][]{abb+18} as a proxy for burst fluence. 
The limitations associated with using \snr as a proxy for burst fluence could be addressed using full voltage baseband data, a possibility that we discuss further in \S\ref{sec:future}.

\subsection{CHIME/FRB Injections}
\label{sec:injections}

The CHIME/FRB injection system characterizes instrumental selection effects and real-time detection biases. While the foundational framework is detailed by \cite{mts+23}, we adopt the latest iteration from \cite{mhj+26}. 
This updated version features a sample size $\sim100$ times larger than its predecessor, expanded specifically to better sample highly scattered bursts and to extend selection-bias constraints from $10\,{\rm ms}$ (\citealt{chimefrbcatalog1}) to $30\,{\rm ms}$ at $600\,{\rm MHz}$.

In this framework, a large sample of synthetic FRBs (called \emph{injections}) is drawn from an initial property distribution and assigned random sky locations. To ensure efficiency, bursts with a low detection probability due to low \snr or weak beam response are filtered out before reaching the pipeline. The remaining bursts are passed through the real-time detection system, which assigns an \snr to each successfully detected event.

\cite{mhj+26} additionally provide a parametric selection function modeled from the injection data. However, we utilize the raw injections directly because this parametric selection function requires true FRB fluences as an input, whereas the catalog fluences are strict lower limits for which the function is not well-posed.

\subsection{Data Selection}
\label{sec:data_selection}

To characterize the FRB population while correcting for selection effects, we must use a sample of bursts with robustly known statistics. We remove bursts from \catii that fail to meet any of the following selection criteria:
\begin{enumerate}
    \item Events detected during periods of non-nominal telescope operation (e.g., pre-commissioning or days with elevated RMS noise) and low-S/N events first identified by citizen scientists on the Zooniverse platform\footnote[3]{\url{https://www.zooniverse.org/}}. 
    % These events are identified using the \verb|excluded_flag| and \verb|citizen_science_flag| flags, respectively.
    
    \item Bursts located in the far sidelobes of the primary beam, where the telescope response function is highly chromatic and not well constrained \citep{abb+20}. 
    %These events were identified using \verb|sidelobe_flag|.
    
    \item Events having ${\rm DM}< {1.5}\, {\rm max}({\rm DM}_{\rm NE2001}, {\rm DM}_{\rm YMW17})$ that could be ambiguously identified as extragalactic based on either model. 
    %{This choice is consistent with \catone and \cite{smb+23}, intended to reduce the systematics induced by uncertainties in Galactic DM models and the potential misidentification of FRBs due to large localization uncertainties. We relax this criterion in \S\ref{sec:access_dmmw} to assess its impact on our results.}
    
    \item Bursts with \verb|bonsai| ${\rm S/N}\leq12$ are excluded to avoid potential human identification errors during the verification process, which could otherwise affect the completeness of our low-S/N sample. 
    
    \item Bursts with a dispersion measure too low (${\rm DM}<100~\dmunits$) to be confidently distinguished from Galactic sources. 
    
    \item Bursts with scattering timescales exceeding $30\,{\rm ms}$ at $600\,{\rm MHz}$, as inferred from injections; beyond this limit, telescope selection effects are not well constrained \citep{mhj+26}. 
    
    \item Events with ${\rm DM}_{\rm NE2001}>100~\dmunits$ to minimize the variance in our results involved with the Milky Way contribution to the DM, as we fix $\overline{\mathrm{DM}}_{\mathrm{MW}}$ to $80~\dmunits$ (see \S\ref{sec:pop_model}). This cut ensures that events falling within $10^\circ$ of the Galactic plane are removed from the sample.
    
    \item All subsequent bursts from known repeating sources, retaining only the first burst from each repeater. This approach addresses several challenges. First, we cannot definitively know which sources in our sample will eventually be identified as repeaters; treating the first detection as an ``apparent one-off'' thus provides a consistent baseline. Second, because known repeaters are predominantly identified at lower redshifts due to observational bias, excluding them entirely would remove roughly $2\%$ of our closest sources and skew our results toward higher redshifts. Finally, while repeaters continue to show distinct morphologies compared to apparent one-offs \citep{pgk+21, csp+25, scm+25}, filtering for specific temporal widths or bandwidths would introduce further complications due to the telescope's complex and chromatic beam response. 
\end{enumerate}

To ensure consistency, the relevant physical cuts are applied to the injection sample as well.
As described by \cite{chimefrbcatalog1} and S23, the \snr cut is the dominant factor, accounting for roughly 50\% of the excluded bursts. Ultimately, {\nfrbsample} bursts in \catii survive all selection criteria, compared to 225 in S23.

%The following Section details the model used to infer the underlying FRB properties using the catalog and injection samples.

\section{FRB Population Model}
\label{sec:pop_model}

% The framework to simultaneously model the volumetric rates, energy function and DM budget of FRBs was developed by \citet{jpm+22b}. However, the sample of FRBs in \catii lack redshift estimates, hence as in S23, we model the observed FRB rate as a joint distribution in fluence ($F$) and dispersion measure (DM), 
% \begin{equation}
%     R(F, \mathrm{DM}\,|\,\boldsymbol{\lambda}) = \int dz \, R(F, z)\, P(\mathrm{DM}\,|\,z).
% \label{eq:rate_distribution}
% \end{equation}
% Here, the joint distribution $R(F,z)$ defines the production rate of FRBs per unit fluence at a given redshift, assuming that the underlying energy function is redshift-independent; as described in \S\ref{sec:catalog}, fluences presented in \catii are strict lower-limits and we use \snr derived from the real-time detection pipeline as a proxy. The conditional probability $P({\rm DM}\vert{}z)$ represents the likelihood that an FRB originating at redshift $z$ yields the observed DM. 
% {Notably, this model assumes a single population of extragalactic bursts sharing a common progenitor and/or emission mechanism, which may not be the case.}
% The model is parameterized by $\boldsymbol{\lambda} = \{\Phi_0,\, \gamma,\, E_{\rm char},\, n,\, \alpha,\, \mu_{{\rm host},e},\, \sigma_{{\rm host},e}\}$, with its components summarized below.

The framework to simultaneously model the volumetric rates, energy function and DM budget of FRBs was developed by \citet{jpm+22b}. {In applying this framework, we assume that the FRBs in \catii originate from a single, homogeneous population of extragalactic sources and share a common progenitor and/or emission mechanism.} Since these FRBs lack redshift estimates, we follow S23 and model the observed FRB rate as a joint distribution in fluence ($F$) and dispersion measure (DM),
\begin{equation}
    R(F, \mathrm{DM}\,|\,\boldsymbol{\lambda}) = \int dz \, R(F, z)\, P(\mathrm{DM}\,|\,z).
\label{eq:rate_distribution}
\end{equation}
Here, the joint distribution $R(F,z)$ defines the production rate of FRBs per unit fluence at a given redshift, assuming that the underlying energy function is redshift-independent; as described in \S\ref{sec:catalog}, fluences presented in \catii are strict lower-limits and we use \snr derived from the real-time detection pipeline as a proxy. The conditional probability $P({\rm DM}\vert{}z)$ represents the likelihood that an FRB originating at redshift $z$ yields the observed DM. The model is parameterized by $\boldsymbol{\lambda} = \{\Phi_0,\, \gamma,\, E_{\rm char},\, n,\, \alpha,\, \mu_{{\rm host},e},\, \sigma_{{\rm host},e}\}$, with its components summarized below.

$\Phi_{0}$ represents the volumetric rate at $z=0$ above a fixed {isotropic-equivalent} energy $E_{\rm pivot}$ in units of ${\rm Gpc}^{-3}\,{\rm yr}^{-1}$. We set $E_{\rm pivot} = 10^{39}$~erg, which lies above the minimum physical energy threshold $E_{\rm min}>10^{38.5}$ ruled out at 90\% confidence by \cite{jpm+22b}. 
%{FRBs may not have an Emin (e.g., Nimmo+2021). Maybe the fact that you don't see a correlation between gamma and phi is enough to say that this is not important for your analysis?} 

The cosmic evolution of FRB rate density tracks the cosmic star formation history \citep[$\operatorname{SFR}(z)$;][]{md14_sfh} via the smooth scaling power-law index $n$, such that:
\begin{equation}
    \Phi(z) = \frac{\Phi_0}{1+z} \left( \frac{\operatorname{SFR}(z)}{\operatorname{SFR}(0)} \right)^n .
\label{eq:sfr_power_law}
\end{equation}
Here the factor of $(1+z)^{-1}$ term accounts for cosmological time dilation.

The intrinsic energy distribution is modeled as a Schechter function consisting of a power law at lower energies in addition to a high-energy exponential cut-off \citep[e.g.,][]{lml+20, lpw20L},
\begin{equation}
P(E)dE = \frac{1}{E_\mathrm{char}} \left( \frac{E}{E_\mathrm{char}}\right)^\gamma \exp\left(-\frac{E}{E_\mathrm{char}}\right) dE,
\label{eq:schechter_function}
\end{equation}
defined by the differential power-law index $\gamma$ and the characteristic energy $E_{\rm char}$ \citep{schechter1976}, and normalized above $E_{\rm pivot}$, as detailed in \S\ref{sec:method}. 
% The FRB energy function can be parametrized as a Schechter function consisting of a power law at lower energies in addition to a high-energy exponential cut-off , or a broken power law \citep[e.g.,][]{lp19, cjd+24, koh+24}. The broken power law is also often used to describe the energy functions of repeating FRBs \citep[e.g.,][]{koh+24,och+26}. 

To convert the observed FRB fluence to its intrinsic energy, we assume isotropic emission and apply the scaling
\begin{equation}
E(F) = \frac{4\pi D_L^2(z)}{(1+z)^{2+\alpha}}\, F \Delta\nu ,
\label{eq:energy_to_fluence}
\end{equation} 
where $D_L^2(z)$ is the luminosity distance. This scaling depends on the true spectral index $\alpha $ (such that $F\propto\nu^\alpha$) and the spectral bandwidth. We assume a bandwidth, $\Delta\nu$, of $1~{\rm GHz}$ for all bursts in the catalog and discuss the implications in \S\ref{sec:future_baseband}. Alternatively, the spectra index can be interpreted as a frequency-dependent scaling of the event rate ($\Phi\propto\nu^{\alpha'}$), where the prime notation distinguishes it from the spectral index interpretation $\alpha $; we present the results of this rate interpretation in Table~\ref{tab:mcmc_master_top} and discuss in \S\ref{sec:discussion_spectralindex}.

In this work, we fix the Milky Way contribution to the observed DM (${\rm DM}_{\rm MW}$) to 80~$\dmunits$ for each FRB, comprising of the median MW disk value of $\sim50~\dmunits$ from NE2001 \citep{ne2001} {for high-Galactic-latitude lines of sight} and the fiducial halo value of $\sim30~\dmunits$ \citep{dgbb15}. We assess the impact of fixing the Milky Way contribution in \S\ref{sec:assess_dmmw}. 
Extragalactic DM ($\dmeg$) consists of two components: the cosmic DM ($\dmcosmic$) arising from the diffuse intergalactic medium, and the host galaxy DM ($\dmhost$) arising from the FRB's host galaxy. 
{Here, we use the probability distribution for $\dmcosmic$ as described by \cite{mpm+20}, which accounts for the fluctuations around the mean DM contributions from the IGM due to intersecting halos and large-scale structures; all related parameters are fixed as in \cite{jpm+22b} and S23. 
However, this formulation of $\dmcosmic$ does not capture the distinct contributions of known local foregrounds, including nearby galaxies such as M31 and M33, as well as the Virgo cluster.
%Intervening halos which can contribute $\sim\mathcal{O}(100)~\dmunits$ to the DM budget, varying significantly based on sightline \citep{lak+22,slp+23,sad+26}. 
%This contribution is redshift dependent, affecting high-redshift sources more, and could be absorbed into the inferred host DM, biasing it high along sightlines intersecting dense structures.
}

{Similar to S23 and other population analyses \citep[e.g.,][]{jpm+22b, hjg+25}}, the rest-frame host contribution to the total observed DM, $\dmhost'$, is modeled as an analytical log-normal distribution $P(\dmhost')$, where $\dmhost=\dmhost'/(1+z)$. The distribution is parametrized by $\mu_{{\rm host},e}$ and $\sigma_{{\rm host},e}$, such that $\mathrm{median}(\dmhost') = e^{\mu_{\rm host},e}$ and variance is $(e^{\sigma_{{\rm host},e}^2}-1)\,e^{2\mu_{{\rm host},e}+\sigma_{{\rm host},e}^2}$. Since $\mu_{{\rm host},e}$ and $\sigma_{{\rm host},e}$ are themselves the shape and scale parameters of the log-normal $P(\dmhost')$ distribution, we use the subscripts $e$ and $10$ throughout this work (e.g., $\mu_{{\rm host},10}$ and $\sigma_{{\rm host},10}$) to denote whether these are expressed in natural- or common-log space, respectively.
We also explore the existence and impact of a potential redshift evolution of the host DM value following \cite{llu26} in \S\ref{sec:discussion_host}.

\section{Methodology}
\label{sec:method}

The mathematical formalism of this framework is elaborated by S23; a conceptual summary is provided here.

To infer the intrinsic properties of FRBs, we compare the observed \catii sample with simulated populations, corrected for selection effects using CHIME/FRB injections. As described in \S\ref{sec:catalog}, we use \snr from the real-time detection pipeline (\verb|bonsai|) as a proxy for burst fluence. To avoid computationally prohibitive multiple campaigns and system shutdowns, we utilize a single injection campaign spanning a broad property phase space \citep[\S\ref{sec:injections};][]{mhj+26} and transform it into any desired population model using property-dependent weights.
% Simulated bursts with low detection probabilities are filtered out based on \snr and beam response before entering the pipeline, and are treated as non-detected.
Each injected burst is assigned a weight representing the ratio of the model-predicted number of bursts to the initial density of the synthetic population, incorporating the survey duration $(\Delta t_{\rm obs})$, sky fraction $(f_{\rm sky})$, and total simulated injections $(n_{\rm simulated})$. The scattering time $\tau $ and intrinsic width $w$ are drawn from the fiducial distributions determined using the method outlined by \cite{mhj+26}, which assumes that $P(\tau)$ and $P(w)$ are separable and normalized to unity. However, unlike \cite{mhj+26}, we incorporate scattering upper limits in this work, giving the shape ($\eta$) and scale ($\zeta$) of fiducial distributions for $\tau$ and $w$ as {$(\eta_\tau, \zeta_\tau)={(4.86\times10^{-3},2.03)}$} and {$(\eta_w, \zeta_w)={(8.00\times10^{-4},1.12)}$}, respectively.

% The joint $F$--$\mathrm{DM}$ distribution is evaluated on a numerical grid. The joint $F$--$z$ distribution, $R(F,z)$, is evaluated on an intrinsic grid of 500 fluence bins and 100 redshift bins; fluence is logarithmically spaced from $0.2$ to $500,000$~Jy-ms, and redshift spans $0.001$ to $4.0$ with the first $68\%$ of bins logarithmically spaced up to $z=1$ and linear spacing thereafter. The observed DM distribution $P({\rm DM}\vert{}z)$ is evaluated on an intrinsic grid using the same redshift coordinates and 150 linear $\mathrm{DM}$ bins spanning $100$ to $6500~\dmunits$, where $\dmmw$ is subtracted from the observed total to evaluate the extragalactic contribution $\dmeg$. 

The underlying joint $F$–$\mathrm{DM}$ distribution is evaluated on a numerical grid identical to the configuration of S23, using the same redshift, fluence, and $\mathrm{DM}$ binning dimensions. The likelihood is sampled using a logarithmically binned Poisson likelihood evaluated on an \snr--$\mathrm{DM}$ grid consisting of 15 \snr bins and 10 $\mathrm{DM}$ bins, spaced logarithmically over ${\rm S/N}\in[12,200]$ and ${\rm DM}\in[100,3500]~\dmunits$, respectively.

We perform the simultaneous seven-parameter fit via Markov Chain Monte Carlo (MCMC) sampling using the \texttt{emcee} package\footnote[4]{\url{https://emcee.readthedocs.io/en/stable/}} \citep{emcee}. The ensemble utilizes 60 walkers, with 60\% initialized around the Maximum Likelihood Estimate (MLE) and the remaining 40\% drawn from the full prior to ensure broad coverage of the parameter space. Unlike S23, who used a local \texttt{curve\_fit} routine, we determine the MLE using the \texttt{scipy.optimize}\footnote[5]{\url{https://docs.scipy.org/doc/scipy/tutorial/optimize.html}} differential evolution method \citep{scipy}; this approach performs a global, gradient-free search across parameter bounds to navigate complex landscapes without getting trapped in local minima. The process consists of a 5,000-step burn-in and a 25,000-step production run. 
We employ a weighted combination of proposal \texttt{emcee} moves\footnote[6]{\url{https://emcee.readthedocs.io/en/stable/user/moves/}} consisting of 40\% Stretch moves, 40\% Differential Evolution (DE) moves, and 20\% DE Snooker moves. As compared to S23 who relied entirely on standard Stretch moves, incorporating DE and Snooker moves helps the ensemble better handle parameter correlations, escape local minima, and ensures the chains are well-mixed.
Chain convergence is monitored using integrated autocorrelation times ($\tau_{\rm auto-corr}$), effective sample size (ESS), and the Gelman-Rubin $\hat{R}$ statistic \citep{gelman1992}.

\section{Results}
\label{sec:results}

\begin{table*}[t] % Use [t] because starred tables usually only appear at the top of the page
    \centering
    \caption{Best-fit parameters and $68\%$ credible intervals for the $R(F,{\rm DM}|\boldsymbol{\lambda})$ model. The values represent the medians of the posterior distributions, with uncertainties derived from the 16th and 84th percentiles. Descriptions for each parameter are provided in \S\ref{sec:pop_model}.
    \label{tab:mcmc_results}}
    
    \begin{threeparttable}

        \setlength{\tabcolsep}{5pt}
        \begin{tabular*}{0.9\textwidth}{@{\extracolsep{\fill}}lccc}
        \hline \hline
        Parameter & Uniform prior range$^a$ & Best-fit result$^b$ & S23 results \\ \hline
        $\log_{10} \; \Phi_0 \: ^{c,d}$ & [-0.96, 6.43] & $5.06^{+0.15}_{-0.16}$ & $4.9^{+0.3}_{-0.3}$ \\ [0.5em]
        
        $\Phi_0\: ^{d}$ & $\ldots$ & $1.1^{+0.4}_{-0.3} \times 10^{5}$ Gpc$^{-3}$ yr$^{-1}$ & $7^{+9}_{-4} \times 10^4$ Gpc$^{-3}$ yr$^{-1}$ \\[0.5em]
        
        $\gamma$ & [-3.00, 3.00] & $-1.41^{+0.27}_{-0.21}$ & $-1.3^{+0.7}_{-0.4}$ \\[0.5em]
        
        $\log_{10} \; E_{\mathrm{char}} \: ^c$ & [38.0, 49.0] & $41.3^{+0.4}_{-0.3}$ & $41.4^{+0.5}_{-0.5}$ \\[0.5em]
        
        $E_{\mathrm{char}}$ & $\ldots$ & $2.0^{+2.6}_{-1.0} \times 10^{41}$ erg & $2.4^{+5.4}_{-1.6} \times 10^{41}$ erg \\[0.5em]
        
        $n$ & [-2.00, 8.00] & $0.5^{+0.4}_{-0.4}$ & $1.0^{+0.8}_{-0.7}$ \\[0.5em]
        
        $\alpha$ & [-5.00, 5.00] & $-1.0^{+0.5}_{-0.6}$ & $-1.4^{+0.9}_{-1.2}$ \\[0.5em]
        
        $\mu_{\mathrm{host},10}\: ^{c,e}$ & [1.30, 2.70] & $2.30^{+0.07}_{-0.09}$ & $1.93^{+0.26}_{-0.38}$ \\[0.5em]
        
        $\mathrm{Median} \; P(\mathrm{DM}_{\mathrm{host}}')$ & $\ldots$ & $2.0^{+0.3}_{-0.4}\times10^2$ pc cm$^{-3}$ & $0.8^{+0.7}_{-0.5}\times10^2$ pc cm$^{-3}$ \\[0.5em]
        
        $\sigma_{\mathrm{host},10} \: ^{c,e}$ & [0.04, 1.74] & $0.36^{+0.04}_{-0.04}$ & $0.41^{+0.21}_{-0.20}$ \\[0.5em]
        
        $\mathrm{Std. dev.} \; P(\mathrm{DM}_{\mathrm{host}}')$ & $\ldots$ & $2.8^{+0.7}_{-0.9}\times10^2$ pc cm$^{-3}$ & $1.7^{+3.3}_{-1.3}\times10^2$ pc cm$^{-3}$ \\[0.5em]
        \hline
        \end{tabular*}
        \begin{tablenotes}[list]
            \small
            \item[a] All parameter priors are chosen identically to those of S23.
            \item[b] {Tabulated here are the statistical uncertainties resulting from the MCMC framework. We estimate the systematic uncertainties to be $\pm25\%$ in \S\ref{sec:results}, which are  dominant as compared to the statistical uncertainties for $\gamma$, $\mu_{{\rm host},10}$, and $\sigma_{{\rm host},10}$; hence, we utilize the systematic uncertainty values for all subsequent analyses of parameters}.
            \item[c] These parameters are fit in natural log-space, but presented here in common logarithm for ease of interpretation.
            \item[d] The volumetric rate in this work is quoted above an $E_{\rm pivot}=10^{39}$~erg and below a scattering timescale of 30~ms at 600~MHz, compared to the 10~ms scattering time limit used by S23.
            \item[e] $\mu_{{\rm host},e}$ and $\sigma_{{\rm host},e}$ are the log-normal location and scale parameters of the host DM distribution P(${\rm DM}_{\rm host}'$) defined in \S\ref{sec:pop_model}; unlike $\Phi_0$ and $E_{\rm char}$, they are themselves natural-log quantities, so we denote their base-10 values with the subscript 10 (e.g., $\mu_{{\rm host},10}=\mu_{{\rm host},e}\times \log_{10}e$).
        \end{tablenotes}
    \end{threeparttable}
\end{table*}

\begin{figure*}
  \centering
  \includegraphics[width=\linewidth]{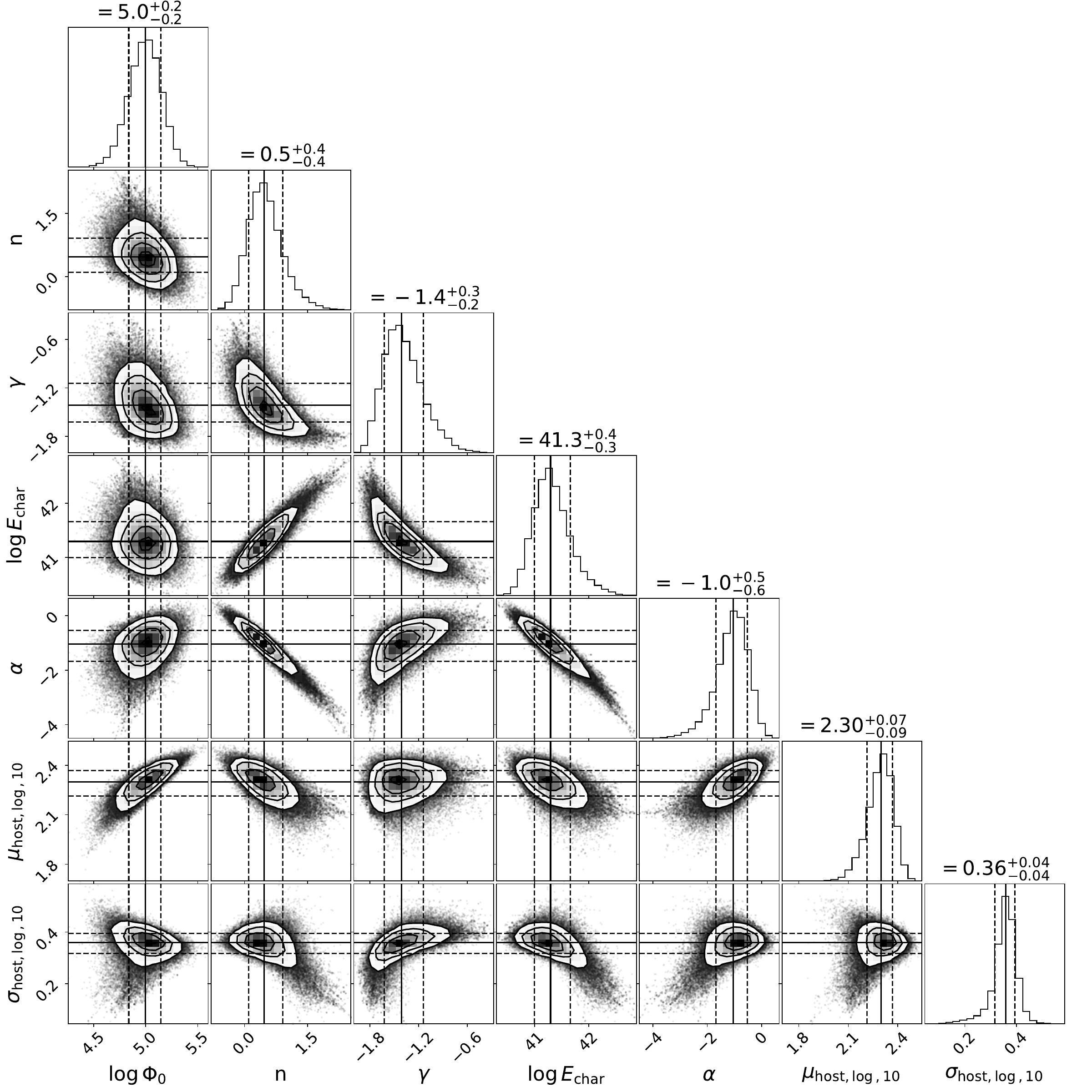}
  \caption{Corner plot of the MCMC posterior distributions corresponding to the $R(F,{\rm DM})$ model described in \S\ref{sec:pop_model}. We apply a thinning factor of 20 for visual clarity. Parameters $\Phi_0$, $E_{\rm char}$, $\mu_{{\rm host},10}$, and $\sigma_{{\rm host},10}$ are fit in natural-log space but presented here in common logarithm for ease of interpretation. The solid vertical lines represent the median of the one-dimensional distributions, while dashed lines enclose the $68\%$ credible intervals. The two-dimensional contours represent the $1\sigma$, $1.5\sigma$, and $2\sigma$ confidence levels. {Systematic uncertainties evaluated in \S\ref{sec:results} are not plotted here.}}
  \label{fig:mcmc_corner}
\end{figure*}

\begin{figure*}
  \centering
  \includegraphics[width=\linewidth]{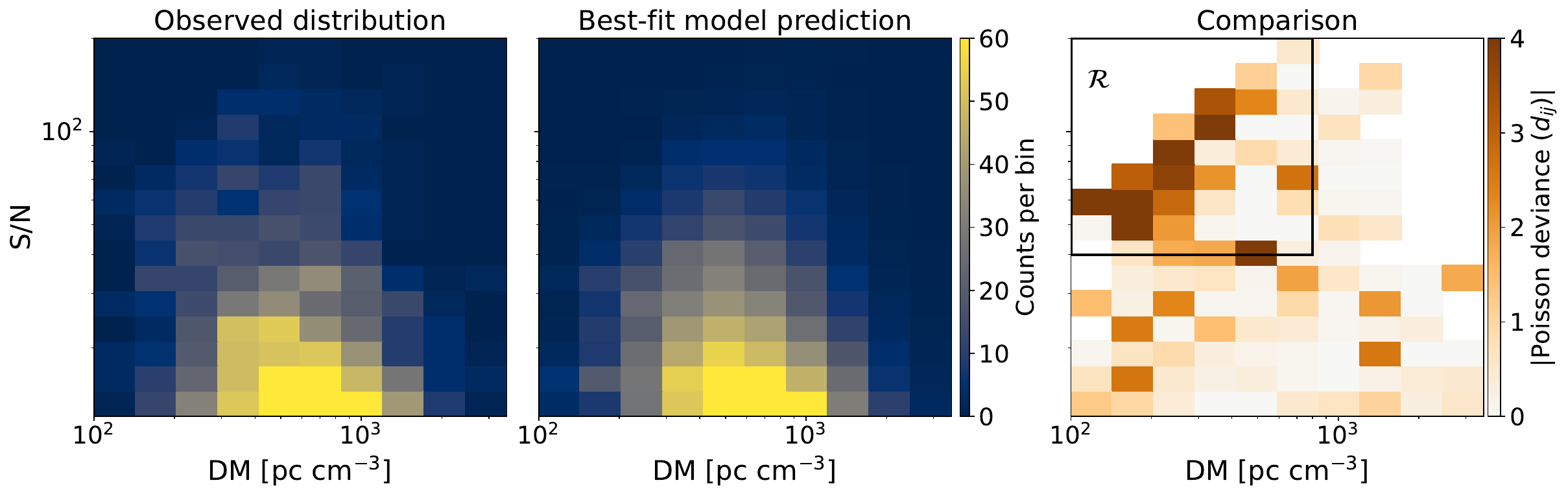}
  \caption{Comparison of the observed $\rm{S/N}\text{--}\rm{DM}$ distribution with the best-fit model predictions. The right panel shows the {absolute values of the} Poisson deviation as defined in Equation~\ref{eq:poisson}. The two distributions are largely consistent within $3\sigma$, with the notable exception of the low-DM, high-\snr regime where the observed counts systematically exceed model predictions. A hint of this feature can be made out visually in Figure~3 of S23; however, it is less apparent due to lower bin counts.}
  \label{fig:catalog_modelprediction_comparison}
\end{figure*}

We use a sample of 1,683 FRBs derived from \catii, along with CHIME/FRB injections to account for instrumental selection effects, to infer the intrinsic properties of the FRB population. The model, described in \S\ref{sec:pop_model}, consists of a joint distribution in fluence and dispersion measure, $R(F, \rm{DM}\,|\,\boldsymbol{\lambda})$, where $\boldsymbol{\lambda }$ represents the free parameters. Following the methodology outlined in \S\ref{sec:method}, we compare the FRB sample from \catii with selection-corrected model populations using a Bayesian Poisson-likelihood fit to obtain robust constraints on these parameters.

All parameters yielded $\hat{R}<1.01$, indicating excellent mixing across chains. Although the energy distribution parameters ($E_{\rm char},\, \gamma$) exhibited higher auto-correlation times ($\tau_{\rm auto-corr}\approx466$), the resulting ESS for all parameters exceeded 1900 with a median ${\rm ESS}\approx2200$, ensuring robust posterior estimates.

The best-fit parameters and the corresponding statistical uncertainties are tabulated in Table~\ref{tab:mcmc_results}, along with the corresponding results from S23 for reference.
{
Following \cite{chimefrbcatalog1}, we evaluate five sources of systematic uncertainties: (i) sensitivity variations over the \catii survey duration, (ii) errors in the beam model, (iii) missing intensity data for some events, (iv) holding the distributions of intrinsic width and scattering timescales fixed, and (v) incompleteness due to human verification. For (iv), we perturb the scale parameters describing the scattering timescale (by $\pm1.0$~ms) and width (by $\pm0.3$~ms) distributions and evaluate the change in the predicted counts, while holding $\boldsymbol{\lambda}$ fixed to the best-fit parameters in Table~\ref{tab:mcmc_results}. Similarly, for (v), we evaluate the change under a raised \snr cut of 14. Adding these in quadrature yields a net systematic uncertainty of $\pm25\%$. The systematic uncertainty is dominant compared to the statistical uncertainties for $\gamma$, $\mu_{{\rm host},10}$, and $\sigma_{{\rm host},10}$; hence, we utilize the systematic uncertainty values for the subsequent analyses.}

Most measured parameters are consistent within uncertainties with those previously reported, though several notable trends emerge: key population parameters are now constrained far better than before, with uncertainties reduced by factors of $\sim2$ in most cases\footnote[7]{{The theoretical expectation of a $\sim2.8\times$ reduction in parameter uncertainty following a $\sim7.5\times$ increase in sample size scales down to a realized improvement of only $\sim2\times$ because of parameter degeneracies and the propagation of variance from using \snr as a proxy for burst fluence.}}.
% and using DM to infer redshift.}}.
Notably, the constraint on the star-formation evolution index $n$ has improved by a factor of $\sim2$, however, it remains statistically consistent with both a non-evolving population ($n=0$) and one tracing star formation ($n=1$) at the $2\sigma$ level.
Figure~\ref{fig:mcmc_corner} displays the posterior distributions from the MCMC analysis as a corner plot. We first evaluate the goodness of fit for the best-fit model in \S\ref{sec:goodness_of_fit}, and we assess the sensitivity of our inference to the choice of $\dmmw=80~\dmunits$ in \S\ref{sec:assess_dmmw}. The parameter constraints and their physical interpretations pertaining to the volumetric rates, energy distribution, and redshift distribution are presented in \S\ref{sec:discussion}.

\subsection{Goodness of fit}
\label{sec:goodness_of_fit}

We evaluate the model's goodness of fit in the same \snr--DM space used to compute the Poisson likelihood during the MCMC fit (\S\ref{sec:method}), since this provides the most direct test of whether the model reproduces the joint distribution the inference was fit to. 
%We use two test statistics: Poisson deviance and Pearson residuals. 

For observed counts $n_{ij}$ and model predictions $\zeta_{ij}$, where $i$ indexes the S/N bin and $j$ indexes the DM bin, the Poisson deviance ($d_{ij}$) is defined as:
\begin{equation}
\label{eq:poisson}
  d_{ij} = -\left[ \zeta_{ij} - n_{ij} + n_{ij} \ln\left(\frac{n_{ij}}{\zeta_{ij}}\right) \right].
\end{equation}
For numerical stability, we set $n_{ij} \log(n_{ij}/\zeta_{ij}) = 0$ when $n_{ij} = 0$. A large {$|d_{ij}|$} suggests that the observation in a specific bin is less consistent with the model prediction.
% We also calculate the Pearson residuals ($r_{ij}$), assuming Poisson variance in the model prediction counts,
% \begin{equation}
% r_{ij} = \frac{n_{ij} - \zeta_{ij}}{\sqrt{\zeta_{ij}}}.
% \end{equation}
% The residual $r_{ij}$ indicates the statistical significance of the deviation in each bin, differentiating between regimes where the observations are in excess ($r_{ij}>0$) or deficit ($r_{ij}<0$) relative to the model. {KS - why change the formalism language?}

Figure~\ref{fig:catalog_modelprediction_comparison} compares the observed FRB distribution with the best-fit $R(F, \rm{DM}\,|\,\boldsymbol{\lambda})$ model, showing the distribution of Poisson deviation across the DM–\snr plane. It shows that in the low-DM, high-\snr regime, the best-fit model deviates by $>3\sigma$ from the observed FRB counts; evaluating Pearson residuals indicates that the model systematically under-predicts counts relative to the observed sample.
Based on this, we define $\mathcal{R}=\{(\rm{DM},\rm{S/N}):\rm{DM}< 800~\dmunits,\rm{S/N}>40\}$ as a region of interest. Although $\mathcal{R}$ contains only {$\sim15\%$} of the total sample counts, it accounts for {$\sim60\%$} of the total Poisson deviance. The boundaries of $\mathcal{R}$ are primarily informed by visual inspection of the deviations in Figure~\ref{fig:catalog_modelprediction_comparison}; however our conclusions are robust to this specific choice. We evaluated an alternative definition for this region, selecting bins where {$|d_{ij}| \geq 3$}, and found that while the specific $p$-values in the following analysis fluctuates, the qualitative results and physical implications remain unchanged.

\begin{figure}
  \centering
  \includegraphics[width=\linewidth]{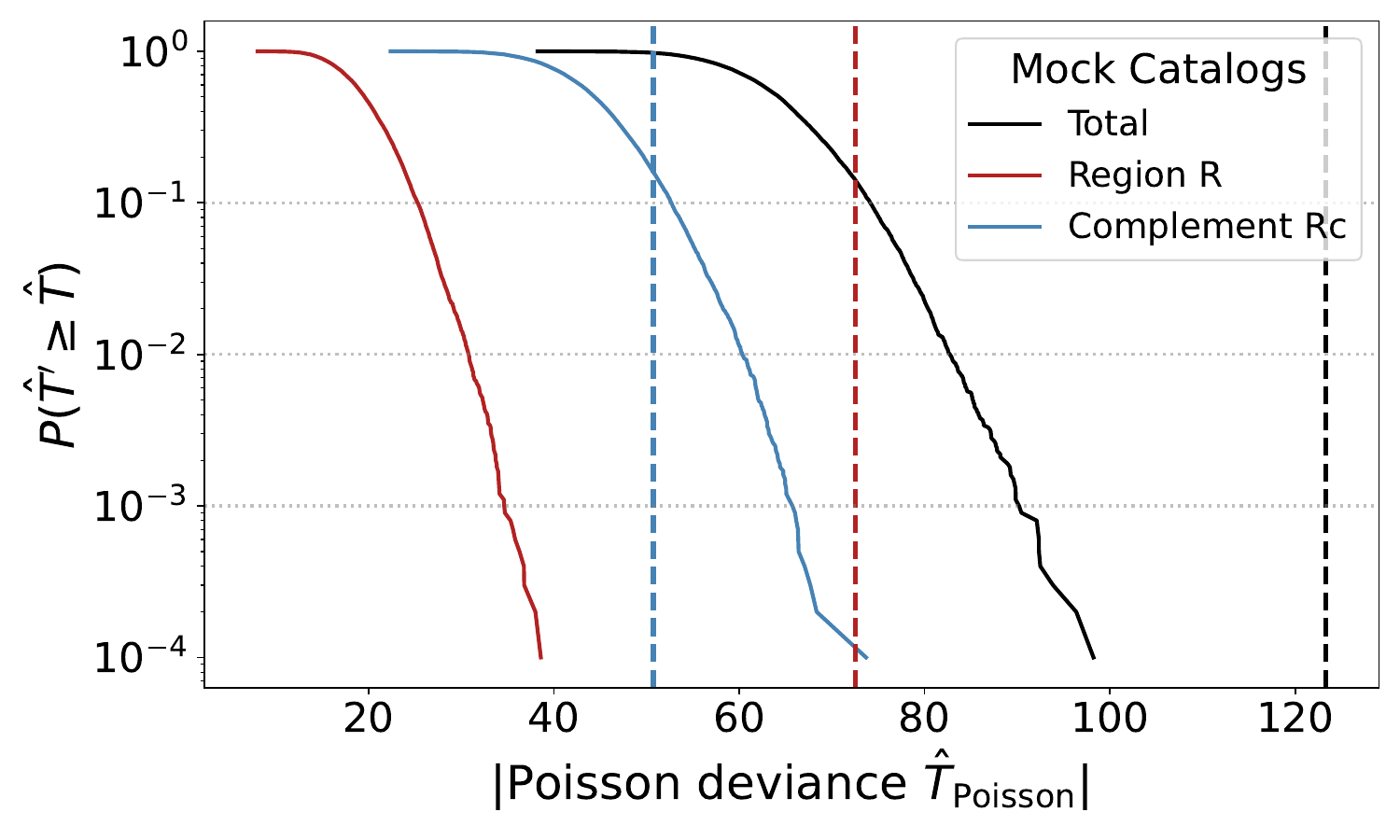}
  \caption{Cumulative survival distributions (solid curves) for the Poisson deviance $T_{\rm Poisson}$. Samples are derived as Monte Carlo realizations of the best-fit $R(F,{\rm DM}|\boldsymbol{\lambda})$ model. The vertical dashed lines indicate the observed statistics from the best-fit model. The statistics are calculated over three domains: the total catalog (black), region {$\mathcal{R}=\{(\rm{DM},\rm{S/N}):\rm{DM}< 800~\dmunits,\rm{S/N}>40\}$} (red), and the complement region (blue). The observed statistics (dashed lines) lie far in the tail of the mock distributions (solid lines) for both the total catalog and region $\mathcal{R}$ ($p < 10^{-4}$), but fall well within the cumulative distribution for the complement region ($p\sim0.2$), demonstrating that the model's global discrepancy is entirely driven by $\mathcal{R}$.}
  \label{fig:mock_catalogs}
\end{figure}

To evaluate the significance of the observed excess, we generate $10^4$ mock observed samples via Monte Carlo sampling from a Poisson distribution with mean $\zeta_{ij}$. For each realization, we calculate the aggregate Poisson deviance ${T}_{\rm Poisson}=\sum_{ij}d_{ij}$ across the full sample, the region $\mathcal{R}$, and its complement $\mathcal{R}_c$. The resulting cumulative survival distribution, $P(\hat{T}'\geq\hat{T})$, is shown in Figure~\ref{fig:mock_catalogs}, where the dashed lines indicate the observed statistics from the best-fit model. The relevant quantity is the intersection of each domain's CDF with the dashed line that defines its $p$-value, noting that the black and red curves must cross their vertical line at $p$-values $<10^{-4}$, outside the range probed and plotted.
The absolute values of $T_{\rm Poisson}$ are not directly comparable across the three domains, as each is a sum over a different number of bins.

For the full observed sample and for $\mathcal{R}$, the observed statistic exceeds all $10^4$ Monte Carlo realizations; the curves therefore truncate at $p=10^{-4}$, the resolution limit set by our sample size, and we quote $p<10^{-4}$ as an upper limit rather than a precise value.
Partitioning the data reveals that this discrepancy is driven by $\mathcal{R}$, whereas $\mathcal{R}_c$ yields {$p\sim0.2$}, indicating a statistically acceptable fit. 
%Similarly, the $T_{\rm Pearson}$ distributions confirm that the full sample exhibits a significant positive residual entirely dominated by the excess in $\mathcal{R}$. The slightly negative residuals in $\mathcal{R}_{c}$, notably above DM $\sim2000~\dmunits$, suggest that the model places excess bursts into this region to maintain consistency with the total observed rate.

We verify these results using $10^4$ bootstrapped realizations of the observed sample from \catii, confirming that the observed excess in region $\mathcal{R}$ is a stable property rather than a stochastic fluke driven by a small subset of events. Additionally, we shuffle the observed sample 10,000 times to test a null hypothesis where the marginal counts in each \snr and DM bin are preserved but the correlation between them is randomized; this further demonstrates that the fitting framework systematically struggles to place enough high-\snr bursts at low DM.

In summary, the $R(F,{\rm DM}|\boldsymbol{\lambda})$ model describes the bulk of the FRB population in $\mathcal{R}_c$ reasonably well, but it does not account for the excess of bright, low-DM bursts in $\mathcal{R}$.
% In the following sections, we examine possible systematic and modeling explanations for this discrepancy.

\subsection{Investigating the Excess of Bright, Low-DM Bursts}

In this section, we examine possible systematic and modeling explanations for the excess of bright, low-DM bursts in $\mathcal{R}$.

{As a first check, we compare bursts inside and outside $\mathcal{R}$ and find no statistically significant differences (two-sample KS test, $p > 0.05$) in sky position (right ascension, declination, Galactic longitude, and Galactic latitude), arrival time, peak frequency, or number of sub-bursts.
However, bursts in $\mathcal{R}$ have shorter durations, and although a larger fraction of them are best fit by a scatter-broadened profile, their scattering times are shorter.
% Their median boxcar width is 2.0 ms, compared with 3.9 ms outside $\mathcal{R}$ (KS $p = 2\times10^{-10}$); 57\% of them are best fit by a scatter-broadened profile, compared with 48\% outside $\mathcal{R}$, and among these scattered bursts the median scattering time is shorter by a factor of $\sim$3 (3.2 vs.\ 10.3 ms; KS $p = 7\times10^{-11}$).
This is expected: at fixed fluence, wider and more scattered bursts are detected at lower S/N, so the S/N threshold defining $\mathcal{R}$ preferentially selects narrower bursts with shorter scattering times.
% The low DMs of bursts in $\mathcal{R}$ may further narrow their measured widths, since dispersion smearing within each frequency channel is smaller at low DM.
Bursts in $\mathcal{R}$ also have broader spectra, with a smaller fraction of them having a spectral FWHM below 150 MHz\footnote[8]{
    {The spectral bandwidths are derived from the ``running'' power-law spectra, $S(\nu) \propto (\nu/\nu_r)^{\alpha + r\ln(\nu/\nu_r)}$. 
    For $r<0$, the spectra peak at $\nu_{\rm pk} = \nu_r e^{-\alpha/(2r)}$,  and are Gaussian in $\ln\nu$ with width $\sigma_{\ln\nu} = (-2r)^{-1/2}$. 
     The half-maximum points lie at $\nu_\pm = \nu_{\rm pk}\, e^{\pm\sqrt{2\ln 2}\,\sigma_{\ln\nu}}$.
    % , giving a spectral FWHM of $\nu_+ - \nu_-$. 
    %We define the in-band bandwidth as the portion of this FWHM that lies within the 400--800 MHz band.
    The 150~MHz cutoff is arbitrary; the difference between bursts inside and outside $\mathcal{R}$ persists for other choices of cutoff.
    While beam effects can artificially narrow the measured bandwidths \citep{apb+23,scm+25}, these measurements reflect the spectral extent seen by the detection pipeline, which is what sets the detection S/N used in this work.
    }
}.
% Specifically, 25\% of bursts in $\mathcal{R}$ have a spectral FWHM below 150 MHz, compared with 41\% outside it (Fisher exact test, $p = 6\times10^{-6}$), and their median in-band bandwidth is 177 MHz, compared with 129 MHz outside $\mathcal{R}$ (KS $p = 3\times10^{-5}$), at similar peak frequencies.
This may reflect their higher S/N, which allows emission to be detected above the noise across more of the band.
}

We also perform jackknife tests by selecting 
(i) events detected at nighttime when radio frequency interference (RFI) activity is reduced and the telescope operates under nominal observing conditions;
(ii) events in different slices of the sky in right ascension (R.A.) to evaluate potential spatial dependencies; and 
(iii) events whose $\sim$arcminute localizations are outside $1\times R_{500}$ of a galaxy cluster with $M_{500} > 10^{14}\, M_\odot$ \citep[identified from the DECaLS photometry by][]{sad+26}, to test FRBs with potentially significant DM contributions from {intervening} clusters as a source of contamination. 
In all cases, the parameter results remain consistent within a $1\sigma$ level relative to those presented in Table~\ref{tab:mcmc_results}. 
{The excess of bright, low-DM bursts in $\mathcal{R}$ is also present in every subset, indicating that it is likely not caused by RFI, a particular region of the sky, or DM contributions from intervening galaxy clusters.}

In Appendix~\ref{app:supplement_results}, we consider modifications to the base model in the form of (i) a delay-time model for the evolution of the rate with redshift (see \S\ref{sec:cosmic_evolution}) and (ii) an energy function described by a broken power law (see \S\ref{sec:discussion_energy}), motivated by their potential to naturally boost the predicted rate of high-\snr bursts at low DMs. 
{In particular, the delay-time model  provides a framework for relaxing our assumption of a single, homogeneous population, by allowing for distinct young and old progenitor channels.}
We note that the {additional parameters in the} extended models {remain unconstrained by the data}, potentially due to the reduced constraining power from using \snr and DM as proxies for burst fluence and redshift, respectively. We defer further investigation of these systematics to future work, discussed in \S\ref{sec:future}.

\subsection{Modeling the Milky Way DM contribution}
\label{sec:assess_dmmw}

\begin{figure*}
  \centering
  \includegraphics[width=\linewidth]{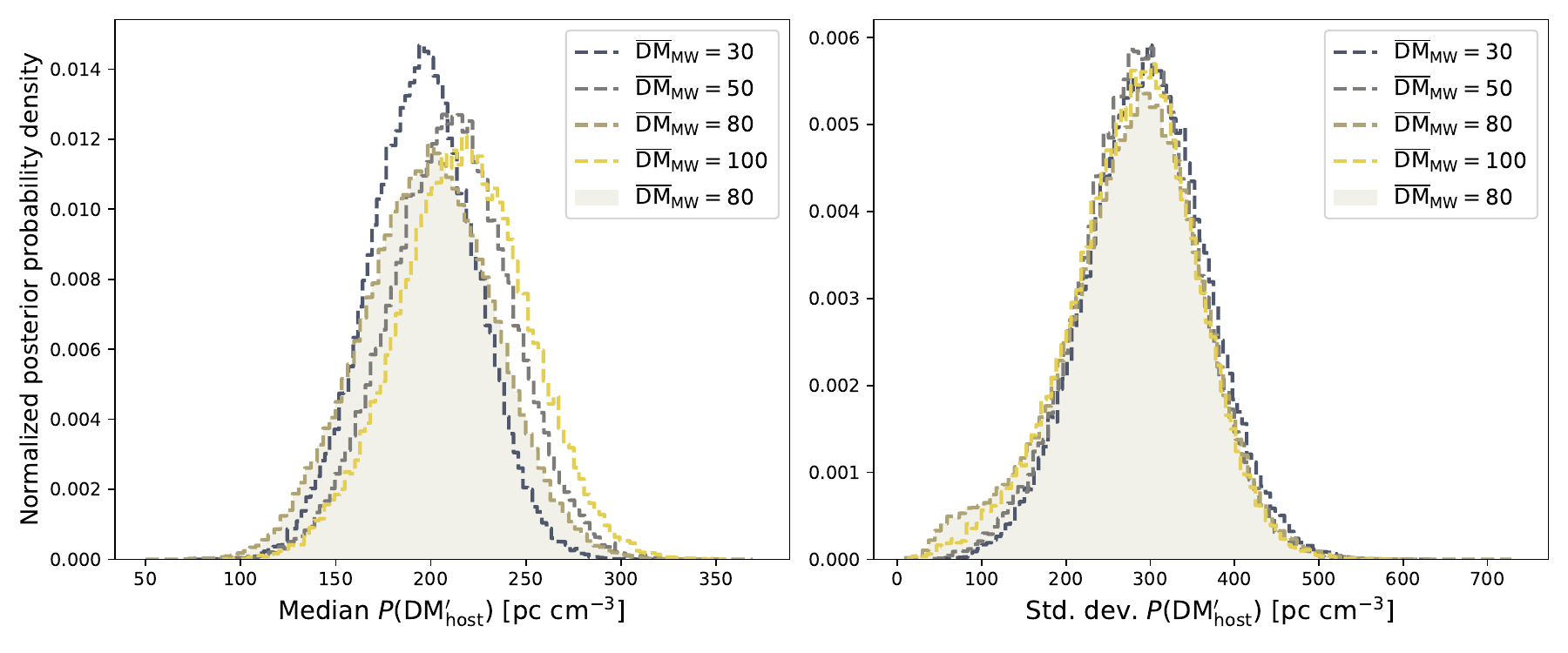}
  \caption{The resulting normalized posterior distributions of median $\dmhost'$ (left) and standard deviation (right) for MCMC runs with $\dmmw$ varied between $30$ and $100~\dmunits$. The best-fit parameters, assuming a log-normal distribution of $\dmhost$, are robust to the choice of $\dmmw$ within this range. }
  \label{fig:vary_dml}
\end{figure*}

We have adopted a fiducial value of $\overline{\rm{DM}}_{\rm{MW}}=80~\dmunits$ in our primary analysis. The assumption of a constant $\dmmw$ is inherently nonphysical, as Galactic interstellar medium (ISM) contributions vary significantly across the sky, as may the contribution of the halo \citep{ne2001, ymw17, kp20, cbg+23, ne2025}. The selection criteria (\S\ref{sec:data_selection}), which exclude events along lines of sight where $\rm{DM}_{\rm{NE2001}} > 100~\dmunits$ are intended to reduce the impact of Galactic variance on our results.

The injections framework is designed to test our bias against the total observed DM \citep{mhj+26}. However, it does not consider how the Galactic DM varies with sky position, and consequently we are blind to the resulting selection effects. For example, if both average Galactic DM and sensitivity change with declination, they could be correlated in a way that is not probed by injections. Thus, we do not consider this systematic in this study. 

We attempt to fit $\dmmw$ as a free parameter along with the base model described in \S\ref{sec:pop_model}, and find the best-fit value to be $\dmmw=45^{+3}_{-2}~\dmunits$. However, in this test, the posterior distributions of $\mu_{{\rm host},10}$ and $\sigma_{{\rm host},10}$ fail to converge and we continue to see a deviation pattern similar to that in Figure~\ref{fig:catalog_modelprediction_comparison}, indicating that we do not have sufficient statistical power to distinguish between the Milky Way and host galaxy contributions to the total DM.
In the absence of robust modeling of Milky Way contribution to the total DM, we examine the sensitivity of our results to the specific value chosen for $\overline{\rm{DM}}_{\rm{MW}}$. We rerun the MCMC analysis while varying the fixed $\dmmw$ between $30$ and $100~\dmunits$. We find no significant shifts in the posterior distributions for the rate and energetics parameters; all best-fit values remain consistent within $1\sigma$. The median and variance of $P(\dmhost')$ are expected to be the most sensitive to this choice; the resulting variation shown in Figure~\ref{fig:vary_dml} is subdominant compared to their statistical uncertainties (Table~\ref{tab:mcmc_results}). 
While a constant $\overline{\rm{DM}}_{\rm{MW}}$ is physically simplistic, our analysis shows the best-fit parameters are not significantly sensitive to the constant value chosen. 
%The resulting variation is subdominant to statistical uncertainties, and the increased variability introduced by more sophisticated Galactic models is not currently supported by the data. 
We therefore maintain our fiducial value of $\dmmw=80\,\rm{pc\,cm^{-3}}$ for the remainder of this work.

\section{Discussion}
\label{sec:discussion}

In \S\ref{sec:goodness_of_fit} we find that the best-fit model fails the goodness-of-fit test. In particular, it underpredicts the rate of high-\snr bursts at low DMs (region highlighted in Figure~\ref{fig:catalog_modelprediction_comparison}), which make up $\sim15\%$ of our observed sample from \catii.
{Several other works have also reported an excess of nearby, low-energy events in Catalog 1 relative to a single-population model \citep{bbt+23,tll+23,zyc+23}, although these analyses do not fully account for CHIME/FRB selection effects or for catalog fluences being lower limits. 
\cite{chimefrbcatalog1} also found a flatter fluence distribution for low-DM events than at high-DM events, implying relatively more high-\snr bursts at low DM, but its model treated $F$ and DM as independent, omitting the coupling expected from their common dependence on redshift.
Our model includes this coupling and, even after calibrating out instrumental selection effects, still underpredicts high-\snr bursts at low DM.
% Catalog 1 also found a flatter cumulative fluence distribution for low-DM events than for high-DM events, 
% % ($\alpha_F\approx -1$ for 100--500~$\dmunits$) than for high-DM events ($\alpha_F\approx -1.8$ above 500~$\dmunits$)
% which implies relatively more high-S/N bursts at low DM. 
% They attributed this to high-DM events preferentially sampling the steeper, high-luminosity end of the luminosity function.
% However, the Catalog 1 model assumed that the $F$--DM distribution factorizes into a power law in $F$ and a free function of DM, and therefore did not include the physical coupling between $F$ and DM expected from their common dependence on redshift.
% Our model includes this coupling, yet still underpredicts the rate of high-\snr bursts at low DM.
}

{Although the model does not capture this excess, it provides a statistically acceptable fit ($p \sim 0.2$) to the remaining $\sim$85\% of the sample. With this caveat in mind, we discuss the resulting constraints on the FRB population in the following subsections.}

\subsection{FRB Volumetric Rate}
\label{sec:discussion_rate}

The parameter $\Phi_0$ represents the volumetric rate of FRBs at $z=0$ above a pivot energy of $E_{\rm pivot}=10^{39}\,{\rm erg}$ and below a scattering timescale of 30~ms at 600~MHz. 
We infer an all-sky volumetric rate of {$1.1^{+0.4}_{-0.3}\times 10^{5}~{\rm Gpc}^{-3}~{\rm yr}^{-1}$}, {where the statistical uncertainties are dominant over the systematic uncertainty of $\pm25\%$}.
%Similar to the case for S23, the statistical uncertainty on the volumetric rate exceeds the systematic uncertainty, and we assume that the same holds for all other parameters, and quote only statistical uncertainties in the following subsections.
Our estimate for $\Phi_0$ is slightly higher but consistent within $1\sigma$ with the measurement of $7.3^{+8.8}_{-3.8} \times 10^4 \, \text{Gpc}^{-3} \, \text{yr}^{-1}$ from S23.
%; this modest increase is likely driven by our extension of scattering timescale cutoff from 10~ms to 30~ms at 600~MHz. Notably, the uncertainties are reduced by factor of $\sim$2. 

These values are broadly consistent with FRB event rate density estimates above $E_{\rm pivot}=10^{39}~{\rm erg}$ utilizing the first CHIME/FRB catalog \citep[e.g.,][]{lwd+25, md2025}.
Because the CHIME/FRB selection function is not well constrained for scattering times exceeding 30~ms at 600~MHz \citep{mhj+26}, we urge caution when comparing our rate to those measured at radio frequencies above $\sim$1~GHz \citep{crt+16, lp19, rav19, lml+20, jgp+22c}. 
Bursts that appear to have low scattering times at higher frequencies can become heavily scattered when observed by CHIME/FRB at 600~MHz, lowering their detection probability. 
Thus, CHIME/FRB may not be sensitive to the same burst population detected at higher frequencies, potentially biasing our volumetric rate estimate low for broadband FRBs. 
Thus, CHIME/FRB may not be sensitive to the same burst population detected at higher frequencies, { in particular, sources embedded in dense, turbulent environments, where scattering is strongest}, and potentially biasing our volumetric rate estimate low for broadband FRBs.
{We test a scenario where the volumetric rate is frequency-dependent (see Appendix~B in S23), $\Phi\sim\Phi(z)\nu^{\alpha'}$, finding mild preference for a negative slope with $\alpha'=-0.7\pm0.4$ (see Table~\ref{tab:mcmc_master_top}), implying FRBs may be rarer at higher frequencies; however at $\lesssim2\sigma$ significance, this result also remains statistically consistent with no frequency dependence in FRB rates.}
%While \cite{lwd+25} replicate the methodology of S23 used here, \cite{md2025} implement a fundamentally different approach. Instead of simultaneously fitting $\Phi_0$ within a marginalized posterior, they determine the volumetric rate by scaling simulation statistics to match the observed count over the survey window. Furthermore, their modeling relies on an analytic beam profile rather than the forward-modeled injection pipeline, altering the treatment of selection effects and fluence lower limits (see \S\ref{sec:catalog}). Thus a direct comparison of our results with \cite{md2025} must be made with caution.
                          
By the volumetric rate above, most cataclysmic progenitor channels, e.g. compact object mergers, do not dominate the local non-repeating FRB population because their rates are orders of magnitude beneath that inferred here \citep{cyz+18, rav19, mb+22, cjd+24, gbk+25}.
However, core-collapse supernovae (CCSNe) are a notable exception: with an event rate density of $\sim10^5~{\rm Gpc}^{-3}~{\rm yr}^{-1}$, they remain compatible with our volumetric rate estimate and so cannot be excluded as a dominant channel. 
%{Our volumetric rate estimate is also consistent with Magnetar (that primarily  originate from ccsne) rate density are favored by the event rate density of FRBs in the local universe ({cite}).}
% Since our volumetric rate can also be interpreted as the rate of FRB sources (because of large systematic uncertainty on sky rate and repeaters being subdominant), 
% , which — within our observing window — are effectively non-repeating. This rate is consistent with a population of distinct, newly-formed magnetars, each contributing at least one detected burst, provided the local magnetar birth rate is sufficiently high {[cite]}. Because the magnetar emission mechanism itself is non-destructive, this scenario does not require a cataclysmic, one-off event per source, distinguishing it from progenitor channels (e.g., compact-object mergers) whose formation and destruction are coincident with the single detected burst.
% However, the FRB emission itself is generally attributed to non-cataclysmic mechanisms — such as magnetar flares, where the source survives to repeat — that are favored by {[specific argument/reference]}; whether this extends to the apparently non-repeating population depends on whether such sources represent a distinct class or the high-energy tail of a shared repeater population (see §\ref{sec:repeaters}).

\subsection{Cosmic evolution of FRB rate density}
\label{sec:cosmic_evolution}

\begin{figure}
  \centering
  \includegraphics[width=\linewidth]{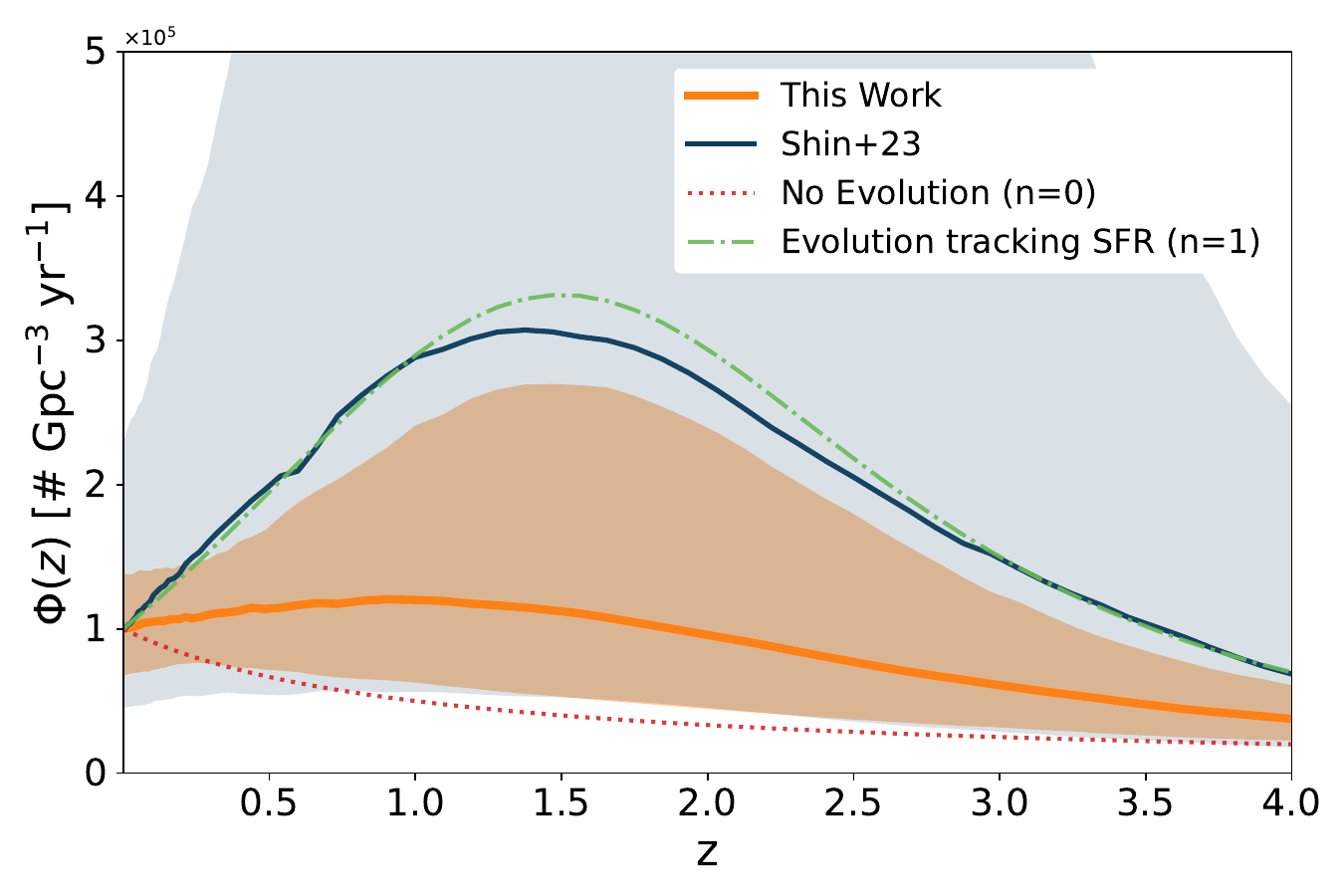}
  \caption{The best-fit results for $\Phi(z)$ from the MCMC run, describing the evolution of FRBs relative to the cosmic star-formation rate (SFR). We compare results obtained from the \catii sample (this work) with those of S23, who evaluated the model using the \cati sample. To facilitate visual comparison,  
  {each $\Phi(z)$ curve is rescaled by a constant factor to shift the curves vertically without changing their shape, so that $\Phi_0 = 1 \times 10^5 \, {\rm Gpc}^{-3} \, {\rm yr}^{-1}$ above $E_{\rm pivot} = 10^{39}~{\rm erg}$.}
  In this work, $\Phi _{0}$ is estimated to be {$1.1^{+0.4}_{-0.3}\times 10^{5}~{\rm Gpc}^{-3}~{\rm yr}^{-1}$}, which is slightly greater than but consistent with the value of $7.3^{+8.8}_{-3.8} \times 10^4 \, \text{Gpc}^{-3} \, \text{yr}^{-1}$ reported by S23. The black dotted line represents $n=0$ (indicating that FRBs do not trace the SFR), while the dash-dotted line corresponds to $n=1$ (indicating that the FRB rate linearly tracks the SFR). Shaded regions denote the $68\%$ confidence intervals.}
  \label{fig:evolution}
\end{figure} 

Modeling the intrinsic volumetric rate density of FRBs helps distinguish between progenitor classes. If FRBs originate from young stellar populations, such as magnetars, their rates should track cosmic SFR and peak near $z\sim2$. Conversely, a peak below $z\lesssim1$ would suggest an origin in older stellar populations or their late-stage by-products. Following previous works \citep[e.g.,][]{jpm+22a}, we model the evolution of FRB rates over time as a power law that traces the cosmic SFR, parameterized by the index $n$; this parameterization enables additional functional freedom in describing the population, rather than encoding a physically motivated model.
Since CHIME's sensitivity restricts the sample to $z<1$, where the star-formation history is rising with redshift, a higher  $n$ increases the mean redshift of the population distribution. To match observed counts, the model compensates by steepening either the spectral index ($\alpha$), which suppresses high-redshift detections via an increased $k$-correction, or the energy function ($\gamma$), which reduces the number of intrinsically energetic bursts. These trade-offs result in the anti-correlations seen in Figure~\ref{fig:mcmc_corner}. 

Our best-fit power-law index, $n=0.5^{+0.4}_{-0.4}$, yields a substantial reduction in uncertainty compared to the $n=1.0 _{-0.7}^{+0.8}$ constraint from S23 (see Figure~\ref{fig:evolution}). {Note that the statistical uncertainties in our best-fit value of $n$ are dominant over the systematic uncertainty of $\pm25\%$.} However, owing to the error margins, this result remains statistically consistent with both a non-evolving rate density ($n=0$) and an evolution tracing the SFR ($n=1$) at the $2\sigma$ level. Similarly, under a rate-interpretation of $\alpha ^{\prime}$ (see {Appendix~\ref{app:supplement_results}}), the alternative best-fit index $n'=0.9^{+0.7}_{-0.6}$ is also compatible with both scenarios. Consequently, the exact evolutionary path remains unconstrained.
%w{This aligns with the directional trend in S23. Curiously, this is opposite to the trend in \cite{jpm+22a}, where the rate interpretation yielded a lower $n$, though all of these results remain consistent within uncertainties.}

{The discrepancy between the data and the best-fit model in Figure~\ref{fig:catalog_modelprediction_comparison} could be explained if FRBs arise from two progenitor channels, associated with young and old stellar populations, each tracing the star formation history with a delay set by its formation mechanism.
FRBs from the old channel would then be concentrated at lower redshifts and, owing to their proximity, appear predominantly at lower DMs and higher \snr.
Our base model cannot produce such a low-redshift peak: its rate scales with the SFR and therefore always peaks at $z \sim 2$, with $n<1$ only flattening the evolution relative to the SFR. 
The delay-time model instead shifts the peak to lower redshift and suppresses the rate at high redshift.
However, as noted in Appendix~\ref{app:supplement_results}, the data do not constrain the delay-time distribution parameters, potentially due to the lack of redshift measurements.
}

We find that the current sample of non-localized FRBs lacks the power to constrain the evolution of FRB rate density. Despite utilizing a sample that is 7.5 times larger than previous implementations, S23 and \cite{lwd+25}, we cannot place significant constraints on either SFR-tracing or delay-time evolutionary models, (see {Appendix~\ref{app:supplement_results}}).
Other studies claiming stronger constraints using the \cati sample often rely on simpler telescope selection models compared to what injections can capture. For instance, \cite{wv24} find evidence for evolution with SFR using CHIME/FRB Catalog 1, but their preferred short delay time is weakly supported and uses an approximate, forward-modeled representation of telescope selection effects. Similarly, \cite{cp25} claim a non-parametric rate that diverges from the SFR, but they correct for CHIME selection effects using a simple flux limit cut that does not account for completeness above that limit.
Additionally, other studies \citep[e.g.,][]{hgc+22, zz2022} claiming that FRBs from \cati do not evolve with SFR differ in their underlying methodology and approximations, such as using total DM to infer $z$ or testing different models rather than explicitly fitting parameters, which can explain the contrasting conclusions.

Applying the $F$--DM framework to a sample of 94 FRBs in \cati with raw-voltage (``baseband'') data \citep{aaa+24}, \cite{gbk+25} find that the majority of sources do not track star formation and exhibit a delay time of $1.9_{-1.3}^{+1.5}$~Gyr. However, it remains unclear how their results are affected by sample size constraints and edge effects in the selection function $P({\rm \snr}\vert\,F, {\rm DM})$. 

FRBs with direct host-galaxy localizations provide exact redshifts and allow us to track evolution directly. 
\cite{jpm+22a} and \cite{jgp+22c} achieved tight constraints on SFR evolution from only a small sample of FRBs, including localized ones, observed at $\sim$GHz frequencies using the Australian Square Kilometre Array Pathfinder (ASKAP) and the Parkes radio telescope.
Both studies find that the FRB population evolves with redshift similarly to the star-formation rate, with \cite{jpm+22a} ruling out a non-evolving population at better than 98\% confidence level. In contrast, based on our analysis using a sample of unlocalized FRBs, the exact evolutionary path of the FRB population remains unconstrained.
%Cross-frequency comparisons between ASKAP ($\sim\,$GHz) and CHIME/FRB ($\sim600$~MHz) samples require careful consideration of the frequency dependence of the FRB rate.

\subsection{FRB Energy Function}
\label{sec:discussion_energy}

\begin{figure}
  \centering
  \includegraphics[width=\linewidth]{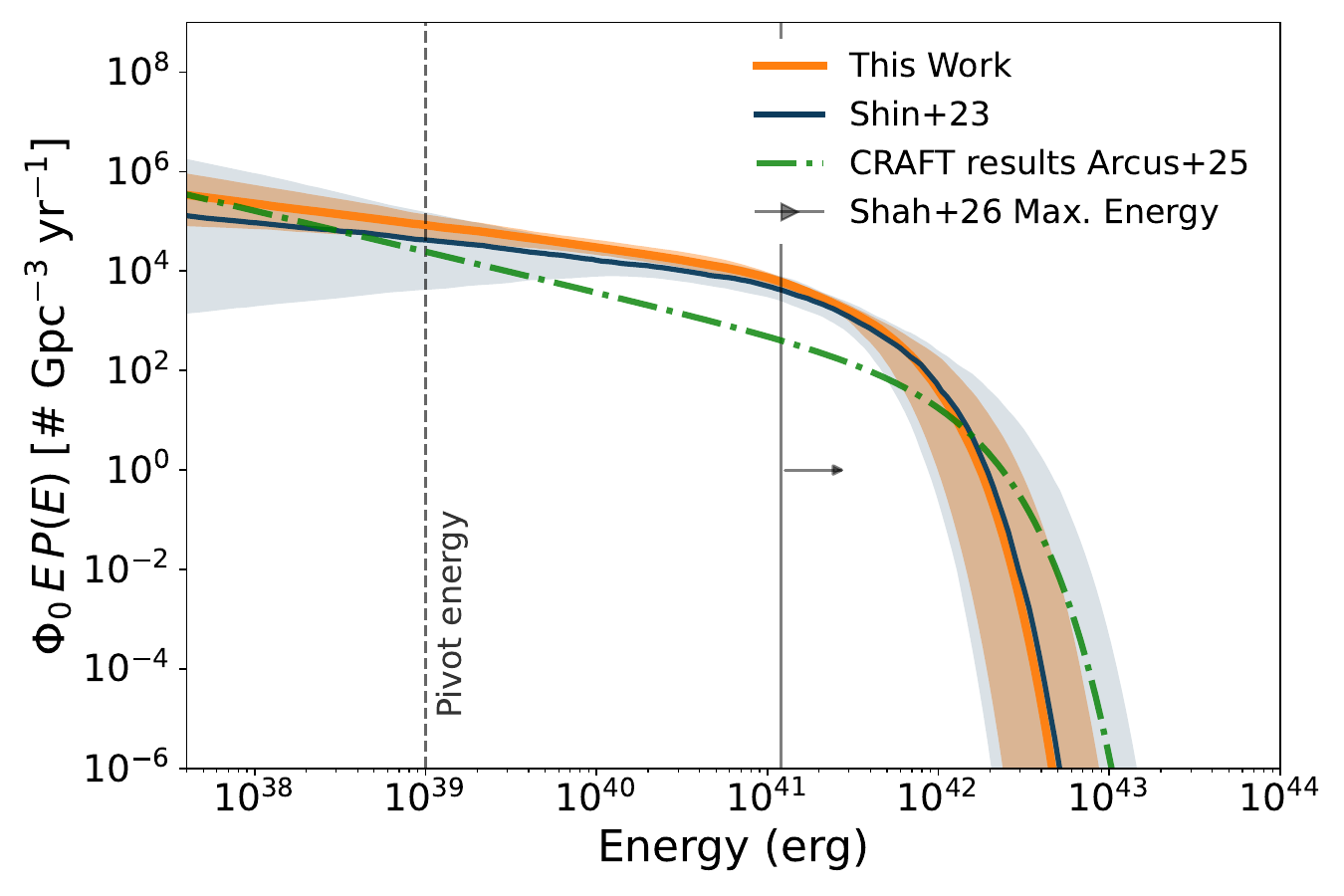}
  \caption{The best-fit Schechter function derived from MCMC simulations. We compare results obtained from the \catii sample (this work) with those of S23, who used the \cati sample. The volumetric rate is reported for energies above $E_{\rm pivot}=10^{39}~{\rm erg}$, indicated by the vertical dashed line. {The solid gray line marks the lower limit constraint on the maximum energy ($E^{\mathrm{max}}_{\mathrm{iso}} = 1.2^{+0.7}_{-1.1}\times10^{42}$~erg) from \cite{shk+26}, reduced by its uncertainty to a conservative value of $1.2\times10^{41}$~erg.} 
  The dash-dotted green line shows the best-fit Schechter energy function using $\gamma=-1.82$ and $E_{\max} \sim 6.3\times10^{41}$~erg from \cite{aje+25}, and assuming a 1~GHz bandwidth and $\log_{10}\Phi_0=4.47$ above $10^{39}$~erg at 1.3~GHz \citep{jgp+22c}. Shaded regions denote the $68\%$ confidence intervals.}
  \label{fig:energy_function}
\end{figure}

We assume an energy function described by a Schechter function and find a differential power-law index of {$\gamma=-1.41^{+0.27}_{-0.21}({\rm stat.})\pm0.35({\rm sys.})$} and an exponential cut-off energy of $E_{\rm char}=2.0^{+2.6}_{-1.0}({\rm stat.})\pm0.5({\rm sys.})\times 10^{41}$~erg. {Since the {systematic} uncertainty in $\gamma$ is dominant, we refer to it for the subsequent discussion.}
We do not test the minimum energy. 
%The pivot energy, $E_{\rm pivot}$, is set to $10^{39}~$erg, which is greater than the minimum physical energy threshold of $E_{\rm min}>10^{38.5}~$erg ruled out at 90\% confidence by \cite{jpm+22b}. 
%Because $P(E)$ is normalized above $E_{\text{pivot}}$, a high $E_{\text{pivot}}$, for instance comparable to $E_{\text{char}}$, introduces a coupling between power-law slope $\gamma$ and $\Phi_0$. By placing $E_{\text{pivot}}$ well below $E_{\text{char}}$, the normalization factor remains insensitive to small shifts in the shape of the energy function. A lack of correlation between $\gamma$ and $\Phi_0$ in Figure~\ref{fig:mcmc_corner} indicates that the choice of $E_{\rm pivot}$ is optimal; consequently, the volumetric rate in \S\ref{sec:discussion_rate} is well-constrained and effectively decoupled from the slope of the energy distribution.
The resulting energy function is plotted in Figure~\ref{fig:energy_function}. While $\gamma$ is slightly steeper compared to the work of S23, our results remain consistent with theirs within uncertainties. Notably, the uncertainties on both parameters are over a factor of two smaller than in the previous measurement.

\cite{shk+26} used \catii to find a lower limit constraint on the maximum energy as $E^{\mathrm{max}}_{\mathrm{iso}} = 1.2^{+0.7}_{-1.1}\times10^{42}$~erg, therefore we don't expect many FRBs in our sample to populate the region above this limit.
Our result is broadly consistent with \cite{gbk+25}, who constrained $\gamma =-1.94^{+0.14}_{-0.12}$ using a sub-sample of CHIME/FRB Catalog 1 with baseband data \citep{aaa+24}, differing at the $\sim2\sigma$ level. Baseband data provides accurate fluence measurements, eliminating the need for using \snr as a proxy for burst fluence. We discuss this further in \S\ref{sec:future_baseband}.

\cite{jpm+22b} used a sample of FRBs observed by the Parkes radio observatory and ASKAP and found a steep power-law slope of $\gamma=-2.09^{+0.14}_{-0.10}$. \cite{hjg+25} reiterated the analysis using a sample of FRBs observed by the DSA, FAST, and CRAFT surveys to constrain a minimum FRB energy of $\log E_{\rm min}=39.47^{+0.54}_{-1.28}$~erg, which is consistent with the value fixed in this analysis. Similarly, \cite{aje+25} used a sample of 63 FRBs observed by ASKAP, including 28 FRBs localized to a host galaxy and found that the FRB energy distribution can be adequately described by both a pure power-law as well as a Schechter function (with $\gamma=-1.82\,\pm\,0.12$). 
The upcoming Canadian Hydrogen Observatory and Radio-transient Detector \citep[CHORD;][]{chord_overview} will perform complementary FRB searches alongside CHIME/FRB; its larger fractional bandwidth (300--1500 MHz) will enable more precise spectral index measurements, helping to constrain the frequency dependence of the FRB energy distribution.

{In our base model, we assume a Schechter energy function. However, the burst energy distributions of a few repeating FRBs is better described by broken power laws \citep{koh+24,och+26}, and for the broader FRB population, it is possible that the energy function instead follows a broken power law with an exponential cutoff. Relative to a Schechter function, such an energy function allows for a higher density of intrinsically low-energy FRBs, which may manifest as an excess of high-\snr bursts detected in the local universe, i.e., at lower DMs, as seen in Figure~\ref{fig:catalog_modelprediction_comparison}. We therefore test a broken power-law energy function in Appendix~\ref{app:supplement_results}.}
% Another possible explanation for the discrepancy seen in Figure~\ref{fig:catalog_modelprediction_comparison} is the form of the energy function: a broken power law {\citep[often used to describe the energy functions of repeating FRBs, e.g.,][]{koh+24,och+26}}, as opposed to a Schechter function, allows for a higher density of intrinsically low-energy FRBs, which may manifest as a set of high-\snr bursts detected in the local universe, i.e., at lower DMs. We describe this model and its implementation in Appendix~\ref{app:supplement_results}.
However, we find that the data do not have the statistical power to distinguish between a broken power law and a Schechter function within the current framework. As a result, the Poisson deviations seen in Figure~\ref{fig:catalog_modelprediction_comparison} for our base model -- which employs an energy function described by a Schechter function -- persist (see Figure~\ref{fig:poisson_result}).

We have not considered evolution of the energy function with redshift in this work. Several works have attempted to model the redshift evolution of FRB luminosities, or their emitted energies \citep[e.g.,][]{cjd+24, zlz+24, zzr+25, cp25, lwd+25}. In particular, \cite{lwd+25} use CHIME/FRB {\cati} along with CHIME/FRB injections \citep{mts+23} and forward model the energy and redshift distributions, finding no evidence for significant redshift evolution in energy. On the other hand, \cite{cjd+24} use the same sample and a non-parametric Lynden-Bell's $C^{-}$ method to conclude that there is strong evolution of FRB energies with redshift, for which FRBs appear intrinsically brighter at higher redshifts. These discrepancies highlight the need for a large sample of localized FRBs with known selection effects in order to probe the energy evolution of FRBs \citep[e.g.,][]{aaa+25}.

\subsection{FRB Spectral Index}
\label{sec:discussion_spectralindex}

The parameter $\alpha$ represents the average FRB spectral index, defined as $F\propto\nu^\alpha$, for which we find a best-fit value of $\alpha=-1.0^{+0.5}_{-0.6}$. {The statistical uncertainties in $\alpha$ are dominant over the systematic uncertainty of $\pm25\%$}. As shown in Figure~\ref{fig:mcmc_corner}, $\alpha$ and $E_{\rm{c}har}$ are anti-correlated; this is expected, as a more positive spectral index makes high redshift bursts appear brighter, lowering the energy required for an observed fluence, which in turn requires a population with a lower $E_{\rm char}$ to maintain consistency with the observed count of high-energy events.

Our best-fit result of $\alpha$ is statistically consistent with that found by S23 ($\alpha = -1.4^{+0.9}_{-1.2}$). Assuming gaussian errors, our constraint on $\alpha$ is inconsistent with zero at $>2\sigma$, suggesting that the typical FRB spectrum is not flat and may have a larger flux at lower radio frequencies. This is in agreement with the $\sim-1$ spectral index that Pleunis et al. (in prep.) found from directly fitting the average spectrum of FRBs using an unbiased sample from \citet{aaa+24}.

Our revised value remains consistent with that estimated at a higher frequency of 1.5 GHz \citep[e.g.,][]{msb+19,jpm+22a}.
Note that the spectral index interpretation as defined herein (see Equation~A4 of S23) treats all FRB spectra as identical power laws; any deviation from a single population-averaged $\alpha$ would propagate as a redshift-dependent bias through the $(1+z)^{2+\alpha}$ term.
Whether this assumption is true is unclear; one could compare with average spectra, for example, of repeater FRB sources. However, it is unclear whether repeaters share a universal spectrum, or whether the general FRB population has the same average spectrum as repeating sources even if repeaters did share one, so comparisons with individual repeater spectra are presently of limited value. 
We note, however, that repeaters, which tend to be narrow-band \citep{pgk+21, csp+25}, constitute only $\sim3\%$ of CHIME detections \cite{abb+25, rn4}, while \citet{scm+25} demonstrate that $>85\%$ of FRBs detected by CHIME appear more broadband in baseband data, which effectively mitigates secondary beam effects.
Repeater burst spectra can vary wildly such that a single power-law index does not provide an adequate general description \citep{hss+19}.
Some authors have attempted to define a synthetic spectrum consisting of that obtained effectively by combining many different bursts from a single repeater \citep[e.g. FRB 20220912A, for which][find an effective spectral index near 1.5 GHz of $-2.6 \pm 0.21$]{zlz+23}.

Another possible quantification of the variation of observability of FRBs with radio frequency is to treat the rate as frequency dependent, that is, examining the number of bursts as a function of their central frequency, such that $\Phi\sim\Phi(z, \nu)= \Phi(z)\nu^{\alpha'}$. When interpreting $\alpha'$ as a frequency-dependent rate, every FRB is effectively treated as a narrow-band burst. While this holds for repeaters, since FRBs observed by CHIME appear more broadband in baseband data \citep{scm+25}, this assumption may therefore not hold in general.
It remains unclear whether the same broadband behavior would be recovered from a repetition of this analysis using baseband data from \catii, or whether these bursts are intrinsically broadband rather than only appearing so within CHIME's observing bandwidth.
Nonetheless, we apply and present this interpretation in Appendix~\ref{app:supplement_results}, where we find $\alpha'=-0.7^{+0.4}_{-0.4}$, statistically consistent with that found by S23, $\alpha'=-1.1^{+0.5}_{-1.0}$, but again, inconsistent with zero at $\sim2\sigma$ confidence.
In this interpretation, independent of spectral extent, the number of bursts is considered as a function of central frequency.
As discussed by \citet{jpm+22a}, the actual frequency dependence is likely some combination of the two views that we have examined; future observatories with broader spectral coverage than CHIME/FRB -- such as CHORD \citep{chord_overview} -- could help clarify the true phenomenology.

\subsection{Host Galaxy DM Distribution}
\label{sec:discussion_host}

\begin{figure}
  \centering
  \includegraphics[width=\linewidth]{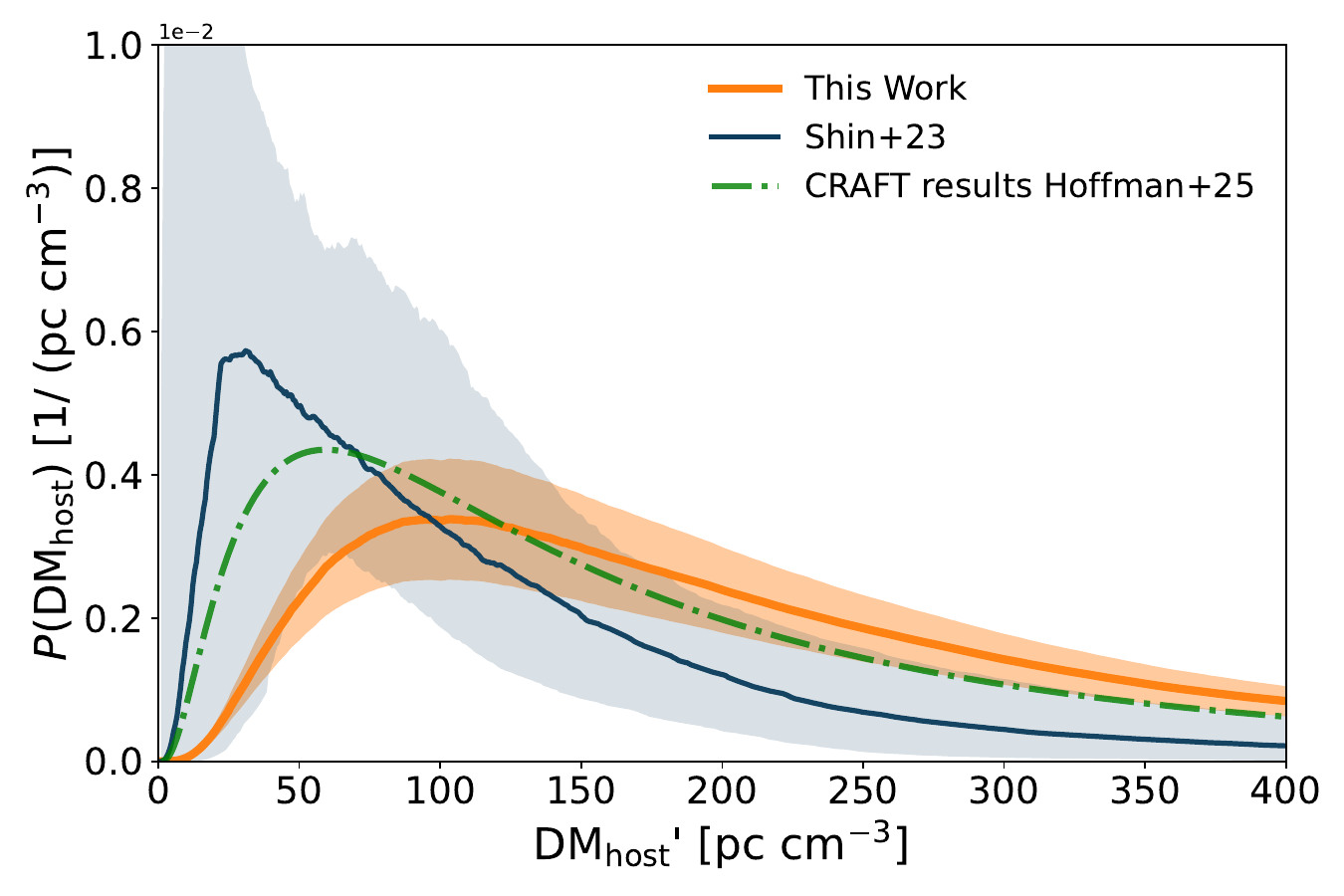}
  \caption{Best-fit host galaxy dispersion measure ($\dmhost$) distribution derived from MCMC simulations. Results obtained from the \catii sample (this work) are compared with those of S23, who utilized the \cati sample. Due to the extended tail of the log-normal distribution, the plot is restricted to $\dmhost'$ values up to $400~\dmunits$. The dashed-dotted green line shows the host distribution defined by best-fit results from \citet{hjg+25}. {The gray shaded region for S23 denotes the $68\%$ confidence intervals, while for this work, the orange shaded region represents a systematic uncertainty of $\pm25\%$.}}
  \label{fig:dm_host}
\end{figure}

We model the rest-frame host galaxy contribution to the total observed DM using a log-normal distribution $P(\dmhost')$, parameterized by $\mu_{{\rm host},10}$ and $\sigma_{{\rm host},10}$ (the subscripts $10$ denote that these are expressed in common-log space), where $\dmhost=\dmhost'/(1+z)$. The best-fit $\mu_{{\rm host},10}$ and $\sigma_{{\rm host},10}$ are presented in Table~\ref{tab:mcmc_results} and correspond to medians and standard deviations of {$199^{+34}_{-36}({\rm stat.})\pm50({\rm sys.})\;\dmunits$ and $285^{+72}_{-86}({\rm stat.})\pm71({\rm sys.})\;\dmunits$}, respectively. As shown in Figure~\ref{fig:mcmc_corner}, $\mu _{\rm{host},10}$ is correlated with $\Phi_{0}$; a larger host DM contribution reduces the mean redshift of the inferred FRB population, where lower intrinsic energies are sufficient for detection, thereby increasing the calculated rate density. By the same logic, $\mu _{\rm{host},10}$ is correlated with $\alpha$, as closer sources require less energy to produce a given observed fluence. 

The posterior distributions of $\mu_{{\rm host},10}$ and $\sigma_{{\rm host},10}$, {even considering systematic uncertainties, are much better constrained as compared to S23, especially on the low-end, resulting in an improvement on the precision by a factor of 2--3}. Our best-fit value is consistent with S23 within $\sim2\sigma$ level. It is consistent with other measurements at the $\sim2\sigma$ level \citep{jpm+22b, crs+25, hjg+25}, albeit higher than typical values of 50--150$~\dmunits$ reported in the literature \citep{smb+18, cc19}. $\mu_{{\rm host},10}$ is correlated with volumetric rate and $\alpha$ (see Figure~\ref{fig:mcmc_corner}), both of which are higher as compared to the results of S23. The value for the standard deviation of P($\dmhost'$) is $60\%$ higher than that of S23 yet its uncertainty range is a factor of $\sim3$ smaller. The resulting $P(\dmhost')$ is still consistent within error bars with that from S23.

{The model described in \S\ref{sec:pop_model} does not consider the evolution of $\dmhost$ with redshift. Following \cite{llu26}, we explore {an evolution in which the median of $P(\dmhost')$ scales as $(1+z)^{n_z}$} in Appendix~\ref{app:supplement_results}, and infer $n_z = 1.6^{+1.4}_{-1.6}$; this value is consistent with zero at the $1\sigma$ level, indicating that a larger sample, preferably including redshift measurements, may be required to reduce the uncertainties associated with $n_z$.}

% {Values compared to localized FRBs and simulations. What does this mean? That there is more DM contribution at low redshifts than we previously thought? More matter in local galaxies? Or is this modeling related?}

% \cite{nal+22} estimates the host-galaxy dispersion measure of FRB 20190520B, which is associated with a dwarf host galaxy at a redshift of $0.241\pm0.001$, to be approximately $903^{+72}_{-111}~\dmunits$
% \cite{cls+24} estimates the host galaxy's DM budget for FRB 20210603A to be $\sim264\pm97~\dmunits$
% ME Orr+24 investigates the contribution of host galaxies to the overall dispersion measures (DMs) for fast radio bursts (FRBs) using the Feedback in Realistic Environments (FIRE-2) cosmological zoom-in simulation suite. At z = 0, they find median host-galaxy DMs of 74 (43) and 210 (94)~$\dmunits$ for older ($\sim$10 Myr) and younger ($\sim$10 Myr) stellar populations, respectively; the value are dependent on the inclination. At $z=1$, the DMs are higher
% Bhandari+23 calculate the host galaxy dispersion measure (DM) contribution for the nonrepeating FRB 20210117A to be approximately 460~$\dmunits$, which is among the highest for ASKAP-localized FRBs.
% JF Mo+23 analyzes the DM contribution from the interstellar medium and circumgalactic medium in host galaxies (DMhost) using simulations from the Illustris and IllustrisTNG projects, noting the differences between young and old populations, especially at $z=0$. Which is interesting...

\subsection{FRB Redshift Distribution}
\label{sec:discussion_redshift}

\begin{figure}
  \centering
  \includegraphics[width=\linewidth]{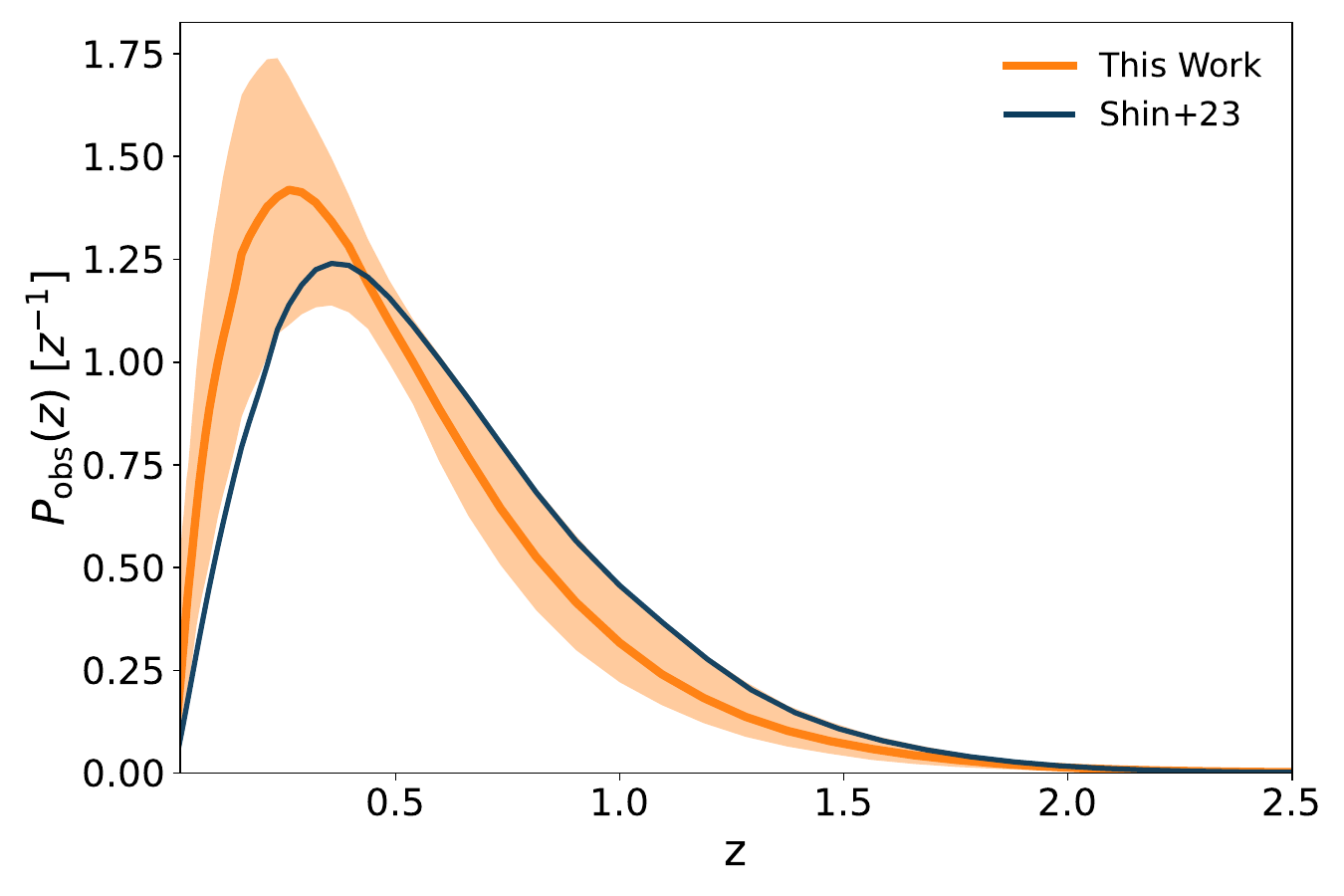}
  \caption{The best-fit $P_{\rm obs}(z)$ derived from the MCMC run. We compare results obtained from the \catii sample (this work) with those of S23, who used the \cati sample. Shaded region denotes the $95\%$ confidence intervals. \\
  }
  \label{fig:pobs_z}
\end{figure}

Figure~\ref{fig:pobs_z} plots the prediction for the redshift distribution of FRBs observed by CHIME/FRB, $P_{\rm obs}(z)$, derived using the methodology described in \S$7$ of S23, who presented $P_{\rm obs}(z)$ as a forecast for redshift distributions of FRBs localized by CHIME/FRB Outriggers. It is the total rate of bursts across space, corrected for cosmological volume and time dilation. 
% The $P_{\rm obs}(z)$ predicted by the best-fit model (in Table~\ref{tab:mcmc_results}) peaks at {$z\approx0.25$}, slightly lower than that of S23. 
{
The $P_{\rm obs}(z)$ predicted by the best-fit model (Table~\ref{tab:mcmc_results}) peaks at $z_{\rm peak} = 0.27^{+0.12}_{-0.08}$ (95\% credible interval), slightly below but consistent with the S23 peak at $z \approx 0.36$.
Because DM is used as a proxy for redshift, redshift and $\mathrm{DM}'_{\rm host}$ are partially degenerate: for the same observed DM, a burst at lower redshift requires a larger host contribution, with a decrease of $\sim0.1$ in $z$ corresponding to an increase of $\sim120~\dmunits$ in $\mathrm{DM}'_{\rm host}$.
The difference between our results and those of S23 lies along this direction, with a lower $z_{\rm peak}$ and higher median $\mathrm{DM}'_{\rm host}$, so the population could have somewhat lower host DMs and higher redshifts than our best fit indicates.}
As discussed by S23, it remains that the bulk of the FRBs detected by CHIME/FRB are expected to originate at $z<2$, limiting their utility as high-$z$ cosmological probes. 
% Using the core CHIME site and the KKO outrigger \citep{lanman_kko}, \cite{aaa+25} provide a sample of 81 FRBs localized to $2''\times\sim60''$ accuracy. Out of the 81 FRBs, only 21 have secure host galaxy associations by the Probabilistic Association of Transients to their Hosts \citep[PATH;][]{aba+21} algorithm, 19 of which have spectroscopic measurements. The mean redshift of this sample is $z\sim0.12$, with the most distant FRB originating at $z\sim0.32$, which is a much narrower range than in our prediction.
A sample of 19 FRBs from \cite{aaa+25}, localized using the core CHIME site and KKO outrigger \citep{lanman_kko} and with spectroscopic redshift measurements, has a mean redshift of $z\sim0.12$ and a most distant event at $z\sim0.32$ --- a much narrower range than in our prediction, {though that sample is further subject to optical selection biases at the high-$z$ and/or faint host galaxy end}.

\begin{figure*}
    \centering
    \includegraphics[width=\linewidth]{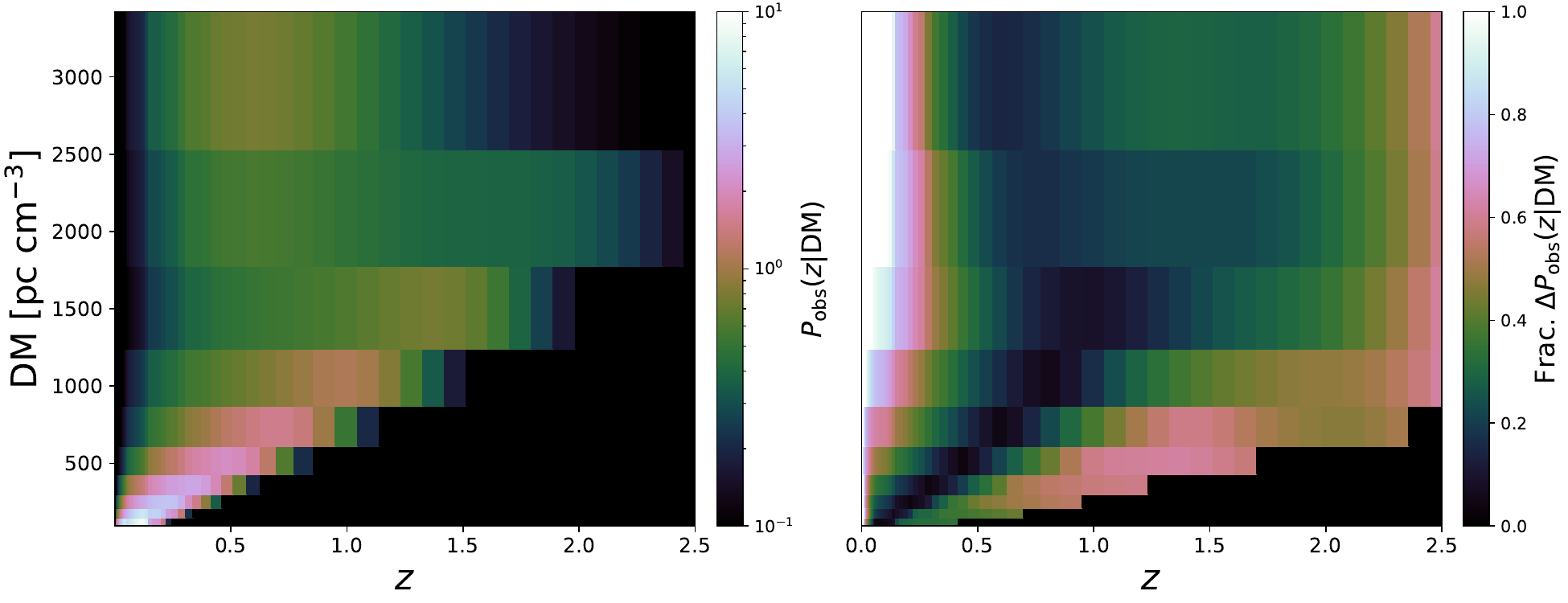}
    \caption{The median (\emph{left}) and fractional uncertainty (\emph{right}) of the $P_{\rm obs}(z|{\rm DM})$ distribution, computed from the MCMC chains thinned by a factor of 2000. A turnover of the DM--$z$ relation remains visible at high DM and low redshift, though the majority of events are located at DM~$\lesssim2000$~pc~cm$^{-3}$ and $z\lesssim2$ (see Figures~\ref{fig:catalog_modelprediction_comparison} and \ref{fig:pobs_z}). The fractional uncertainties in the right panel represent an improvement compared to those reported in S23, notably for ${\rm DM}\gtrsim1500~\dmunits$ across the full sampled redshift range ($z\gtrsim0.3$), as well as at $z\gtrsim2.0$ regardless of DM.}
    \label{fig:Pobsz_givenDM}
\end{figure*}

We present the observed joint redshift and DM distribution, $P_{\rm obs}(z|{\rm DM})$, in Figure~\ref{fig:Pobsz_givenDM}, derived using the methodology described in \S{7} of S23. We continue to observe a turnover at high DMs, {previously observed by \cite{jpm+22b} and S23}, which suggests that a high-DM (${\rm DM}\gtrsim2000~\dmunits$) burst is more likely to originate at a lower redshift ($z\lesssim1$) with a large $\dmhost$ contribution, rather than at high redshift ($z>2$) with a dominant ${\rm DM}_{\rm IGM}$ contribution; this behavior emphasizes the importance of direct FRB redshift measurements in cosmological studies. Additionally, we illustrate the fractional uncertainties associated with $P_{\rm obs}(z|{\rm DM})$ across our MCMC chains. These uncertainties show a marked improvement over S23. The most substantial improvement occurs at ${\rm DM}\gtrsim1500~\dmunits$ across the full sampled redshift range ($z\gtrsim0.3$), as well as at $z\gtrsim2.0$ regardless of DM, where S23's uncertainties approach unity while ours remain moderate. Conversely, in the regions where ${\rm DM}\lesssim1000~\dmunits$ for $0.5<z<1.5$, and at $z\gtrsim2.3$, $P_{\rm obs}(z|{\rm DM})$ is negligible with moderate scatter driven by variations in large-scale structure noise (as previously noted by S23).

In this framework we use DM as a proxy for the redshifts. However, there are several limitations that affect how the DM budget is apportioned across its various contributions, and consequently how distant the burst can be inferred to be -- we use a simplistic model for the Milky Way contribution to the total DM, do not consider evolution of host DM with redshift, and do not consider dense intervening structures along various sightlines.
Despite these limitations, modeling the FRB redshift distribution using non-localized FRBs remains valuable, provided it is paired with more accurate models for the distribution of ionized material in the universe.
Alternatively, as discussed in \S\ref{sec:future_redshifts}, baseband data may provide independent probes of redshift.

\section{Considerations for Future Population Modeling}
\label{sec:future}

In the subsequent subsections, we discuss the current limitations inherent in both the model and the available data, the mitigation of which remains beyond the scope of this study. In this work, we use \snr and DM as proxies for burst fluence and redshifts, respectively; baseband data would provide more accurate fluences (\S\ref{sec:future_baseband}), while outriggers would provide direct redshift measurements (\S\ref{sec:future_redshifts}). Together, these direct measurements would enable us to independently model multiple underlying populations, which remains unfeasible with the current data. Other improvements include considering repetition (\S\ref{sec:future_repeaters}), improved control of selection biases through more comprehensive injection campaigns, and broader frequency coverage enabled by CHORD (\S\ref{sec:other_improvements}).

\subsection{Baseband Data and Burst Morphologies}
\label{sec:future_baseband}

The FRB sample from \catii utilized in this work consists of total intensity data. Because fluence measurements derived from total intensity data represent strict lower limits \citep{apb+23}, we use the \verb|bonsai| \snr as a proxy for burst fluence within our sample. While \verb|bonsai| \snr is useful for calibrating out detection biases, it is not ideal to model FRB fluences. The mapping between burst \snr and fluence is calibrated by injections, marginalized over CHIME's chromatic beam response. This marginalization introduces noise that could be excised exactly given true fluence measurements. 

A subset of FRBs detected by CHIME/FRB is accompanied by raw-voltage (``baseband'') data. By beamforming at the position of the burst on the sky, baseband data enable precise localization and consequently provide accurate fluence measurements \citep{aaa+24}, thereby eliminating the systematics introduced by using \snr as a proxy for burst fluence. The baseband data provide a time resolution of $2.56~\mu\text{s}$ compared to $\sim 1$ ms for total intensity data, enabling robust measurements of burst morphology, including scattering timescales that may otherwise be unresolved \citep[e.g.,][]{scm+25}. 

The forthcoming CHIME/FRB Baseband Catalog 2 {(in prep.)} is expected to increase the number of FRBs with high-fidelity baseband data by an order of magnitude. However, any modeling framework utilizing baseband data must account for additional non-trivial selection effects that are not captured by current injection campaigns. For example, successful baseband capture requires a real-time \snr$>12$ and an accurate real-time DM estimate. Furthermore, downstream baseband localization relies on additional criteria, e.g., whether there is sufficient signal bandwidth to ensure a stable phase solution.

\subsection{Redshifts}
\label{sec:future_redshifts}

To date, we have $\mathcal{O}(100)$ FRBs with host galaxy localizations from all FRB surveys combined \citep{jbc+23,lsr+24,aaa+25a}, and modeling these FRBs as a single population, given the wide range of frequencies at which these surveys operate, requires careful consideration of the instrumental beam response and selection effects \citep{hjg+25}.
CHIME/Outrigger system is projected to provide milli-arcsecond localizations for $\mathcal{O}(100)$ FRBs per year \citep{mlc+21,aaa+25}, using three CHIME-like telescopes deployed at continental baselines to form a very-long-baseline interferometry (VLBI) network. Obtaining a redshift for an Outrigger-localized FRB requires (i) sufficient signal to cross-correlate across sites, (ii) a confident host association using PATH \citep{aba+21}, and (iii) a spectroscopic or photometric redshift for that host; each of these steps is a potential source of selection biases \citep[e.g. see ][on the reliability of PATH associations]{ap+26} that must be accounted for in any population study using this sample. 

Complementary to a sample of localized FRBs with known redshifts, careful modeling of width and scattering is important when performing FRB population modeling \citep{jhp+26} as they enable expansion of the likelihood estimation to a four-dimensional space including fluence, DM, burst width, and scattering timescale \citep[e.g.,][]{wv24}.
Because scattering is dominated by dense turbulent plasma near the source \citep{cls+24, gbj+25}, it can serve as an additional probe of host environments and can help mitigate the ambiguity in the host DM vs the full extragalactic DM contribution \citep{coc22}. 
However, because the $\tau$--DM relation has significant scatter due to fluctuations in DM from other intervening media \citep{ckr+22,occ_2022,mj+26}, caution must be exercised \citep[e.g., no relationship between $\tau_{\rm obs}$ and DM was seen by][]{sdb+25}.
The intrinsic width of an FRB, which scales with redshift as $w_i \propto (1+z)$ \citep{Zhang2023_physics}, offers an additional, independent probe for constraining redshift.
The sample presented in the forthcoming CHIME/FRB Baseband Catalog 2 (in prep.) is roughly $20$ times larger than the current volume of FRBs with known redshifts, offering an opportunity to leverage the statistical power of these additional proxies in future population studies.

\subsection{Considering Repeating FRBs} 
\label{sec:future_repeaters}

If all FRBs are inherently repeating, a population analysis must simultaneously model repetition rates alongside rate density, cosmic evolution, the energy function, and host galaxy properties; \cite{jam23}, and later \cite{rn4} with an expanded number of repeaters, test a small, discrete set of candidate repetition-rate distributions and find that a single population of repeating FRBs can explain the observed population of non-repeaters. Alternatively, if repeaters represent a distinct population, they must be modeled separately from non-repeaters.
However, we lack the ability to predict which sources will repeat in the future, and filtering for specific temporal widths or bandwidths to identify potential repeaters would introduce further complications due to CHIME/FRB's complex and chromatic beam response \citep{pgk+21}. Furthermore, the current injection campaign does not constrain selection biases against repeater-like burst morphologies, limiting our ability to model the two populations independently -- an issue to be addressed by an update to the injection campaign.

Given these limitations, in this work we only use the first bursts detected from known repeaters and model the population of observed FRB sources. A naive exclusion of all known repeaters \citep{rn4} from the sample does not alter our results to an extent greater than the current statistical uncertainties.

\subsection{Other improvements} 
\label{sec:other_improvements}

Operating in parallel, CHIME/Slow is a dedicated backend {to detect radio transients with durations of 10s of millisecond to several seconds} using the same incoming data stream as CHIME/FRB \citep{mlb+26}; it will provide useful insights into CHIME/FRB's bias against temporally wide FRBs \citep{mhj+26}. 
    
Additionally, CHORD will provide a frequency coverage from 300 to 1500~MHz and higher sensitivity as compared to CHIME/FRB \citep{chord_overview}, serving as a powerful engine for discovering high-$z$ bursts. The synergy between the CHIME/FRB and CHORD search pipelines can provide useful insights into FRB spectral properties, mitigating key limitations in current population studies.

% We emphasize that any population analysis leveraging these next-generation datasets must rigorously account for respective (and often non-trivial) selection effects \citep{connor19}. 

\section{Conclusions}
\label{sec:conclusion}

In this work, we have extended the population modeling framework of S23 by utilizing a sample of 1,683 FRBs from the second CHIME/FRB catalog \citep{abb+25}, representing nearly an eightfold increase in sample size. Instrumental selection effects are accounted for using the CHIME/FRB injection framework of \citet{mhj+26}, which characterizes completeness up to scattering timescales of 30~ms at 600~MHz, compared to the 10~ms limit of \citet{chimefrbcatalog1}. This larger sample, combined with a more thorough injection-based selection function, yields at least a twofold improvement in the precision of key parameters describing the observed CHIME/FRB population. However, we note a discrepancy between the best-fit model predictions and the observed sample, wherein the model underpredicts the occurrence of high \snr bursts at low-DMs. 
A qualitative summary of our results, {with the systematic uncertainty highlighted where it is dominant,} is provided below:
\begin{enumerate}
    \item The inferred volumetric rate of FRBs above $E_{\rm pivot}=10^{39}$~erg at $z=0$ is found to be {$1.1^{+0.4}_{-0.3}\times 10^{5}~{\rm Gpc}^{-3}~{\rm yr}^{-1}$} This value is consistent with the event rate densities of core-collapse supernovae (CCSNe) $\mathcal{O}(10^5)~{\rm Gpc}^{-3}~{\rm yr}^{-1}$, which remain a possible dominant channel. 
    
    \item Although the precision on the power-law index governing the evolution of the FRB rate density has improved by a factor of $\sim2$, our constraint of $0.5^{+0.4}_{-0.4}$ remains fully compatible at the $2\sigma$ confidence level with both a non-evolving scenario ($n=0$) and an evolution tracing the cosmic star formation history ($n=1$).

    \item The best-fit energy function exhibits an exponential suppression above $2.0^{+2.6}_{-1.0} \times 10^{41}$~erg and a differential power-law slope of {$\gamma=-1.41^{+0.27}_{-0.21}({\rm stat.})\pm0.35({\rm sys.})$}, which is slightly steeper than the value reported by S23 but shallower than constraints derived from higher-frequency surveys ($\gamma \sim -2.0$). 
    
    \item Both the spectral-index and rate-interpretation frameworks suggest that the average FRB spectrum is not flat, favoring a spectral index of $\alpha = -1.0^{+0.5}_{-0.6}$ and $\alpha' = -0.7^{+0.4}_{-0.4}$. This implies that the FRB population is brighter at lower frequencies.
    
    \item The host galaxy contribution to the total dispersion measure is found to be elevated as compared to the results of S23, with best-fit median and standard deviation of the log-normal distribution $P(\dmhost')$ of {$199^{+34}_{-36}({\rm stat.})\pm50({\rm sys.})\;\dmunits$ and $285^{+72}_{-86}({\rm stat.})\pm71({\rm sys.})\;\dmunits$} respectively. Our best-fit value of the median $P(\dmhost')$ remains consistent with other works \citep{jpm+22b,crs+25}. 
    %We explore a host DM redshift evolution of the form $(1+z)^{n_z}$ and find $n_z=1.6^{+1.4}_{-1.6}$. This matches the results of \citet{llu26} derived from independent instruments and methodologies, supporting a redshift evolution at the $1\sigma$ level; all remaining parameter values are consistent with our base model within $1\sigma$.

    \item The inferred redshift distribution of the observed population peaks at $z \approx 0.27$, which limits the utility of this current sample as a probe of high-$z$ cosmological parameters. 
\end{enumerate}

We are entering an era in which a large number of FRBs possess well-measured properties and are localized with milli-arcsecond precision to secure host galaxy redshifts. Consequently, population modeling can not only refine current parameter measurements, but also allows us to address additional fundamental questions, in particular, whether there are multiple underlying FRB populations described by distinct properties.

\section{Acknowledgments}

We acknowledge that CHIME is located on the traditional, ancestral, and unceded territory of the Syilx/Okanagan people. We are grateful to the staff of the Dominion Radio Astrophysical Observatory, which is operated by the National Research Council of Canada. CHIME operations are funded by a grant from the NSERC Alliance Program and by support from McGill University, University of British Columbia, and University of Toronto. CHIME was funded by a grant from the Canada Foundation for Innovation (CFI) 2012 Leading Edge Fund (Project 31170) and by contributions from the provinces of British Columbia, Québec and Ontario. The CHIME/FRB Project was funded by a grant from the CFI 2015 Innovation Fund (Project 33213) and by contributions from the provinces of British Columbia and Québec, and by the Dunlap Institute for Astronomy and Astrophysics at the University of Toronto. Additional support was provided by the Canadian Institute for Advanced Research (CIFAR), the Trottier Space Institute at McGill University, and the University of British Columbia. The CHIME/FRB baseband recording system is funded in part by a CFI John R. Evans Leaders Fund award to IHS.

The AstroFlash research group at McGill University, University of Amsterdam, ASTRON, and JIVE is supported by: a Canada Excellence Research Chair in Transient Astrophysics (CERC-2022-00009); an Advanced Grant from the European Research Council (ERC) under the European Union's Horizon 2020 research and innovation programme (`EuroFlash'; Grant agreement No. 101098079); an NWO-Vici grant (`AstroFlash'; VI.C.192.045); an NSERC Discovery Grant (RGPIN-2025-06681); an ERC Starting Grant (`EnviroFlash'; Grant agreement No. 101223057); and an NWO-Veni grant (VI.Veni.222.295).

A.P.C. is a Canadian SKA Scientist and is funded by the Government of Canada / est financé par le gouvernement du Canada.
V.M.K. holds the Lorne Trottier Chair in Astrophysics \& Cosmology, a Distinguished James McGill Professorship, and receives support from an NSERC Discovery grant (RGPIN 228738-13). 
K.W.M. is supported by NSF Grant No. 2510771.
D.M. acknowledges support from the French government under the France 2030 investment plan, as part of the Initiative d'Excellence d'Aix-Marseille Universit\'e -- A*MIDEX (AMX-23-CEI-088).
A.P. is a Trottier Space Institute Postdoctoral Fellow.
The AstroFlash research group at McGill University, University of Amsterdam, ASTRON, and JIVE is supported by: a Canada Excellence Research Chair in Transient Astrophysics (CERC-2022-00009); an Advanced Grant from the European Research Council (ERC) under the European Union’s Horizon 2020 research and innovation programme (`EuroFlash’; Grant agreement No. 101098079); an NWO-Vici grant (`AstroFlash’; VI.C.192.045); an NSERC Discovery Grant (RGPIN-2025-06681); an ERC Starting Grant (`EnviroFlash’; Grant agreement No. 101223057); and an NWO-Veni grant (VI.Veni.222.295).
M.W.S. acknowledges support from the French government under the France 2030 investment plan, as part of the Initiative d’Excellence d’Aix-Marseille Université - AMIDEX (AMX-23-CEI- 088).
P.S. acknowledges the support of an NSERC Discovery Grant (RGPIN-2024-06266).
H.M.T. is a Trottier Space Institute Postdoctoral researcher.
A.B.P.~acknowledges support from NASA through the NASA Hubble Fellowship grant \mbox{HST-HF2-51584.001-A} awarded by the Space Telescope Science Institute, which is operated by the Association of Universities for Research in Astronomy, Inc., for NASA, under contract \mbox{NAS5-26555}. A.B.P.~also acknowledges prior support from a Banting Fellowship, a McGill Space Institute~(MSI) Fellowship, and a Fonds de Recherche du \mbox{Qu\'ebec -- Nature} et Technologies~(FRQNT) Postdoctoral Fellowship.
K.N. acknowledges support by NASA through the NASA Hubble Fellowship grant \# HST-HF2-51582.001-A awarded by the Space Telescope Science Institute, which is operated by the Association of Universities for Research in Astronomy, Incorporated, under NASA contract NAS5-26555.
V.S. is supported by a Fonds de Recherche du Quebec—Nature et Technologies (FRQNT) Doctoral Research Award.
A.Pd. acknowledges support from the French government under the France 2030 investment plan, as part of the Initiative d’Excellence d’Aix-Marseille Université - AMIDEX (AMX-23-CEI- 088).
C. L. acknowledges support from the Miller Institute for Basic Research at UC Berkeley.
FRB research at UBC is funded by an NSERC Discovery Grant and by the Canadian Institute for Advanced Research.
J.M.P. acknowledges the support of an NSERC Discovery Grant (RGPIN-2023-05373).
M.D. is supported by a CRC Chair, NSERC Discovery Grant, and CIFAR.

\bibliography{frbrefs}

\appendix

\counterwithin{figure}{section}
\counterwithin{table}{section}
\renewcommand{\thefigure}{\thesection\arabic{figure}}
\renewcommand{\thetable}{\thesection\arabic{table}}

\section{Supplemental Results for Extended Population Model}
\label{app:supplement_results}

We repeat the analysis as described in \S\ref{sec:method} while extending the base model described in \S\ref{sec:pop_model} with the features described below. The results are presented in Table~\ref{tab:mcmc_master_top}.

\begin{itemize}
    \item Under the rate interpretation, $\alpha'$ is interpreted as the exponent characterizing the frequency dependence of the rate, such that $\Phi(z,\nu) = \Phi(z)\nu^{\alpha'}$. As a consequence, the frequency dependencies of fluence and cosmic evolution of FRB rates become coupled -- the detectability of high-redshift FRBs increases as the intrinsic rate is boosted at the rest-frame frequencies that are redshifted into the CHIME's observing band. For a detailed description see Appendix~B in S23.

    \item  
    A general delay-time model with $N$ distinct progenitor channels can be written as
    \begin{equation}
          \Phi(z)=\frac{\Phi_0}{1+z}\sum_{i=1}^{N} f_i\,\frac{I_i(z)}{I_i(0)},\qquad
          I_i(z)=\int_0^{\tau_{\rm max}(z)} d\tau\,p_i(\tau)\,{\rm SFR}\!\left(z_a[t_L(z)+\tau]\right),
    \label{eq:sfr_delay_time}
    \end{equation}
    {where $t_L(z)$ is the look-back time to redshift $z$ and $z_a$ is its inverse.
    % , i.e. the redshift at a given look-back time, so a progenitor with delay $\tau$ formed at look-back time $t_L(z)+\tau$.
    Each channel $i$ has weight $f_i$, the fraction of the local ($z=0$) FRB rate it contributes, with $\sum_i f_i = 1$, so that $\Phi(0)=\Phi_0$ is the local volumetric rate.
    Each channel has a lognormal delay-time distribution (DTD),
    \begin{equation}
        p_i(\tau)\,d\tau = \frac{1}{\sqrt{2\pi}\,\sigma_i}
        \exp\!\left[-\frac{\bigl(\ln(\tau/{\rm Gyr}) - \mu_i\bigr)^2}{2\sigma_i^2}\right] d\ln\tau,
    \end{equation}
    normalized over $0<\tau<\infty$, where $\mu_i$ and $\sigma_i$ are the mean and standard deviation of $\ln(\tau/{\rm Gyr})$.
    Delays longer than $\tau_{\rm max}(z)=t_L(z_{\rm SF})-t_L(z)$ would require the progenitor to form before the onset of star formation, which we take to be at $z_{\rm SF}=20$, so they do not contribute.
    The factor of $(1+z)^{-1}$ accounts for cosmological time dilation between the source and observer frames, and ${\rm SFR}(z)$ is taken from \cite{md14_sfh}.
    In this formulation, all channels share the base model's energy function and differ only in their DTDs and weights $f_i$.}
    % \begin{equation}
      %   \Phi(z)=\frac{\Phi_0}{1+z}\sum_i f_i \int d\tau_i\, p_i(\tau_i)\,\frac{\rm{SFR}(z_a[t(z)-\tau_i])}{\rm{SFR}(0)},
      % \label{eq:sfr_delay_time}
      % \end{equation}
    % where each index $i$ corresponds to a separate evolutionary channel, with corresponding weight within the full population $f_i$ and a delay time distribution (DTD) $p_i(\tau_i)$. $z_a$ is a function that obtains the redshift at a given look-back time. 

    % The simplest version of this formulation is a single evolutionary channel with a delta-function DTD, $p(\tau)=\delta(\tau-\tau_0)$; i.e., all FRBs are assumed to have the same delay time $\tau$. The parameter $\tau$ is unconstrained by the data within the prior range of $[0,14]$~Gyr. This suggests that the data may not have sufficient statistical power to support a more sophisticated and astrophysically motivated version of this model, consisting of two underlying populations described by distinct DTDs and energy functions.

    {We fit the simplest version of this formulation: a single channel with a lognormal DTD, with uniform priors on $\mu_\tau \in [\ln(10^{-5}),\,\ln(14)]$ and $\sigma_\tau \in [0.01,\,3.0]$, corresponding to median delays between 10~kyr and 14~Gyr. Both $\mu_\tau$ and $\sigma_\tau$ remain unconstrained by the data within these ranges (Table~\ref{tab:mcmc_master_top}).} This suggests that the data may not have sufficient statistical power to support a more astrophysically motivated version of this model, consisting of two channels ($N=2$) described by distinct DTDs and, potentially, distinct energy functions.

    \item The double power-law energy function is defined as,
    \begin{equation}
    \label{eq:mod_energy}
      P(E)\propto \left[ \left(\frac{E}{E_{\rm break}}\right)^{\gamma_1} +\left(\frac{E}{E_{\rm break}}\right)^{\gamma_2} \right]^{-1},
    \end{equation}
    where $\gamma_1$, $\gamma_2$ are the power-law indices for the low and high energy regimes, and $E_{\rm break}$ is the break-off energy where the power-law slope changes. 
    This probability distribution function is normalized above $E_{\rm pivot}=10^{39}~$erg.
    % \begin{equation}
    % \label{eq:mod_energy_norm}
    %   P(E\mid E\ge E_{\rm pivot})= \frac{f(E)} {\displaystyle \int_{E_{\rm pivot}}^{\infty} f(E')\,dE'} \qquad\text{for }E\ge E_{\rm pivot},
    % \end{equation} and $P(E)=0$ below $E_{\rm pivot}$. 
    %Unlike the Schechter function, the normalization factor $\int_{E_{\rm pivot}}^{\infty} f(E')\,dE'$ does not reduce to an analytic incomplete gamma function. Therefore, the normalization must be performed numerically. In practice, we define the integration domain as $(E_{\rm pivot}, 10^4\times E_{\rm char})$, as the exponential cutoff renders the tail of the distribution negligible beyond this limit.
    We adopt a parameterization such that $\gamma_2=\gamma_1+\Delta\gamma$; enforcing $\Delta\gamma\in[0,8]$ allows us to test whether the two slopes are meaningfully different. We use a uniform prior on $E_{\rm break}\in[10^{35},10^{49}]~{\rm erg}$. The priors for all other parameters are the same as described in Table~\ref{tab:mcmc_results}. 
    
    While $\Delta\gamma$ remains unconstrained by the data within its prior range, the best estimates of $\gamma$ and $E_{\rm break}$ are consistent with the power-law slope and characteristic cut-off energy constraints for an energy described by the Schechter function (see Table~\ref{tab:mcmc_results}), suggesting that the data lacks the statistical power to constrain the exact form of the energy function.

    \item 
    %Recently, \citet{llu26} reported $1\sigma$ evidence for an increase of host DM with redshift associated with galaxy evolution. By analyzing 94 DSA and CRAFT FRBs with confirmed host redshifts, they found a rest-frame host DM evolution scaling of the form $(1+z)^{n_z}$, with $n_z = 1.62^{+1.48}_{-1.57}$, broadly consistent with simulations \citep[e.g.,][]{zyh+20, mzw+23, obl+24} and observed data analyses (e.g., \citealt{sl25}). \citet{llu26} argued that ignoring a potential evolution may bias inference analyses, given the expected correlations between the evolution and cosmological/astrophysical parameters. Table~\ref{tab:mcmc_master_top} presents our results when allowing for a host DM redshift evolution via the parameter $n_z$, following \citet{llu26}. We infer $n_z = 1.6^{+1.4}_{-1.6}$, very closely matching the value of \citet{llu26} despite our different dataset and instrument characteristics. Our result supports their claim of the existence of a redshift evolution at $1\sigma$. 
    
    {Following \cite{llu26}, we also test a model in which the rest-frame host DM evolves
    with redshift, with median $e^{\mu_{\rm host}}(1+z)^{n_z}$:
    \begin{equation}
    \label{eq:evolve-dm-host}
        P(\dmhost'\,|\,z) = \frac{1}{\dmhost'}\frac{1}{\sigma_{\rm host}\sqrt{2\pi}}
        \exp\left[-\frac{\left(\ln\left[\dmhost'(1+z)^{-n_z}\right]-\mu_{\rm host}\right)^2}{2\sigma_{\rm host}^2}\right],
    \end{equation}
    which reduces to the base model for $n_z=0$.}

    {As in \cite{llu26}, we observe correlations between $n_z$ and $\Phi_0$, $\alpha$ and $\sigma_{\rm host}$, all of which vary within $2\sigma$ error region as compared to the base model. Given the large uncertainties in this evolution, we do not discuss these degeneracies further and leave a more detailed analysis to future work.}
\end{itemize}

\begin{figure*}
    \centering
    \includegraphics[width=0.45\linewidth]{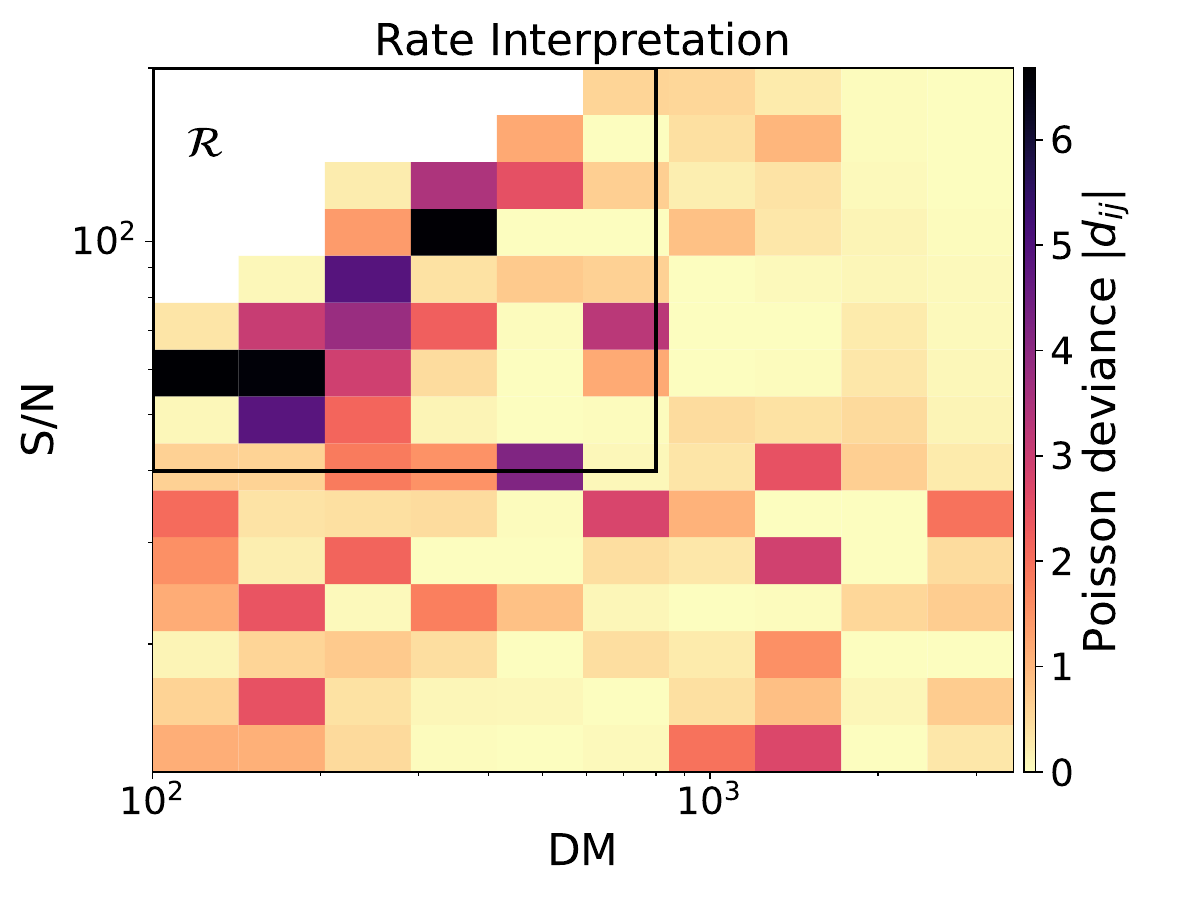}
    \includegraphics[width=0.45\linewidth]{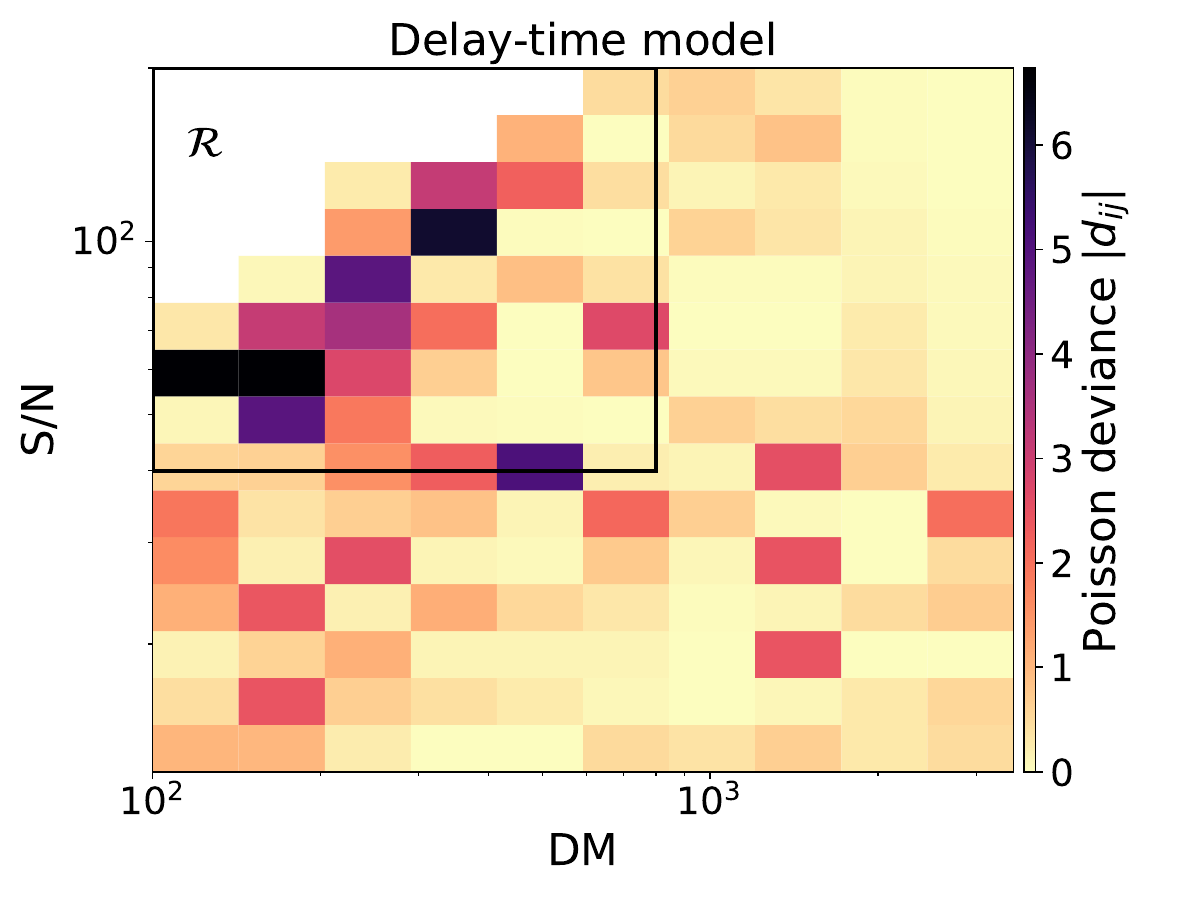}
    \includegraphics[width=0.45\linewidth]{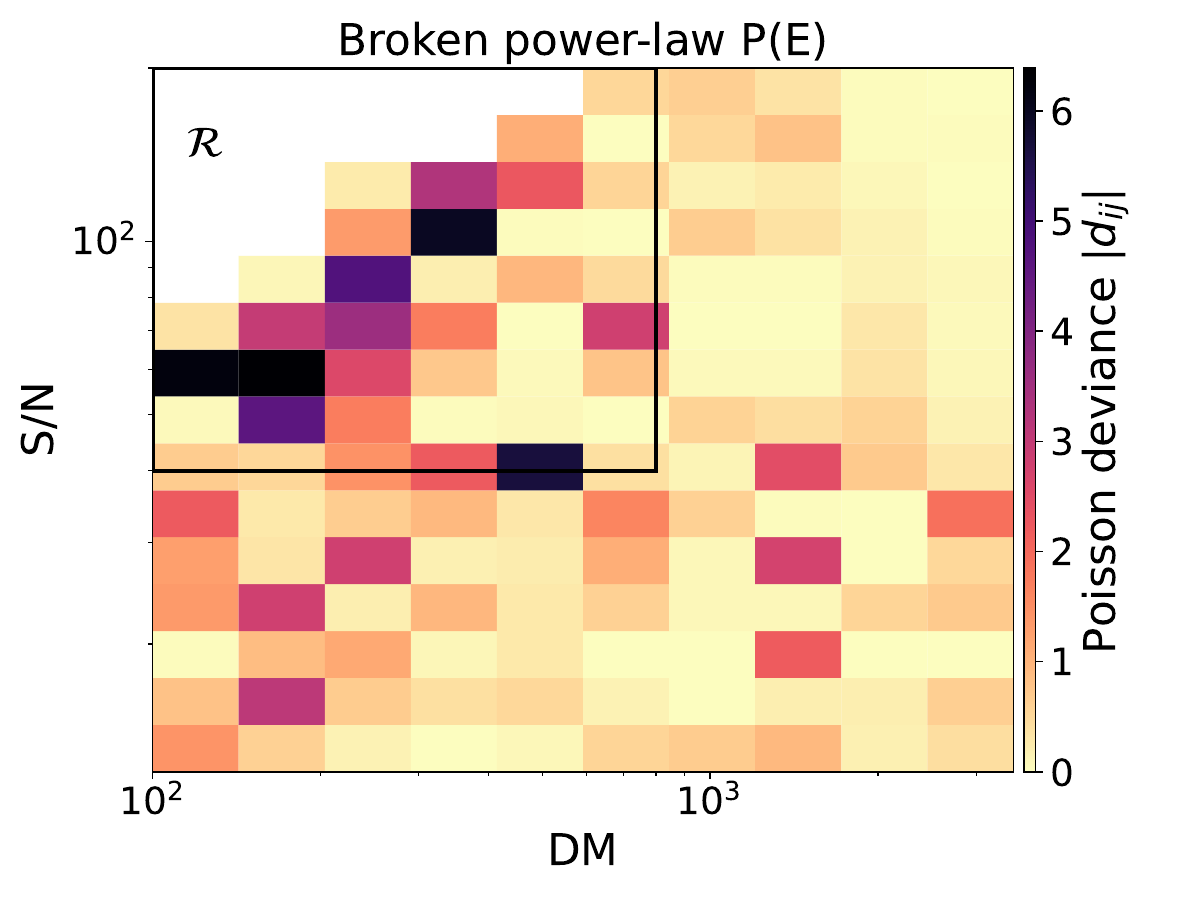}
    \includegraphics[width=0.45\linewidth]{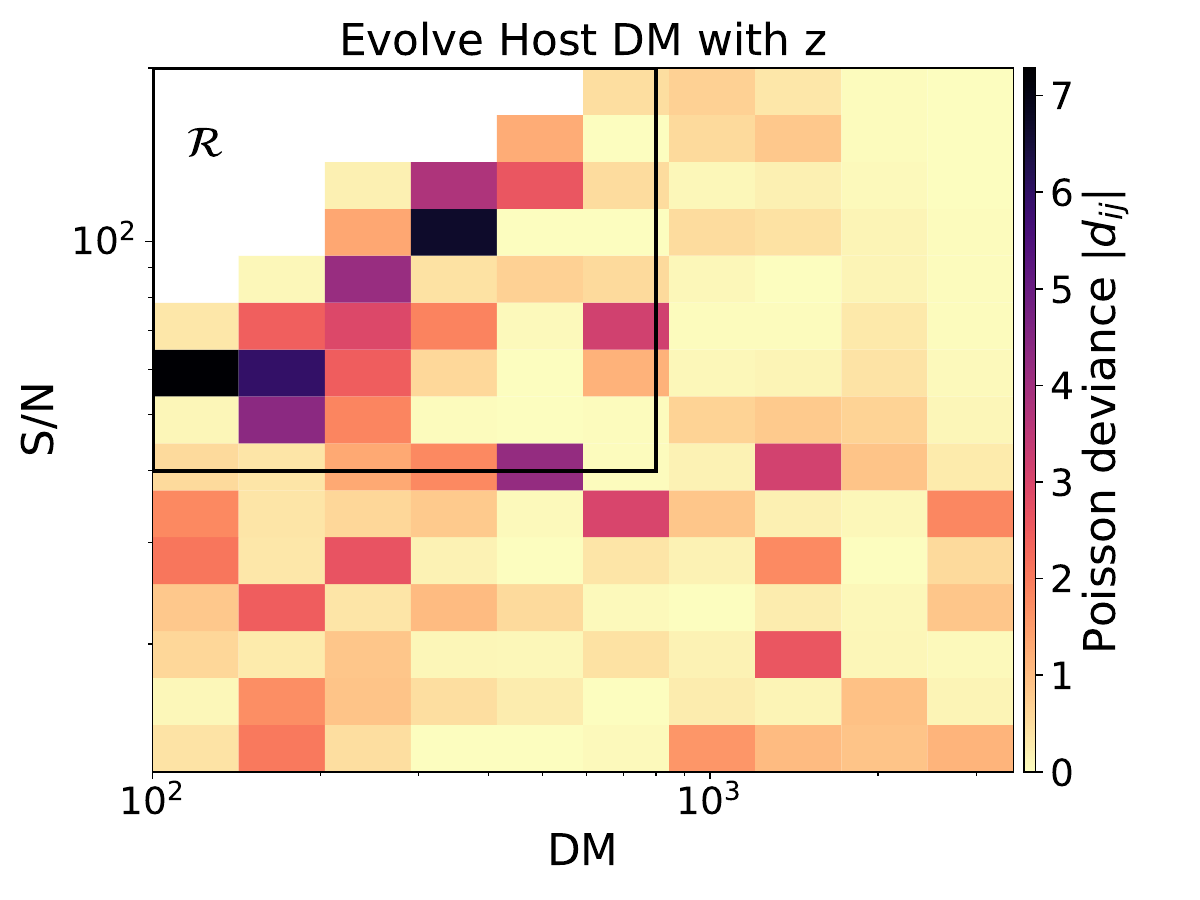}
    \caption{Poisson deviance $|d_{ij}|$, as defined in \S\ref{sec:goodness_of_fit}, between the observed \catii sample and the predicted counts for four modified versions of the base model, described in Appendix~\ref{app:supplement_results}: a rate interpretation of the frequency dependence (\emph{top left}), a delay-time model for the redshift evolution of the rate (\emph{top right}), a broken power-law energy function (\emph{bottom left}), and a model allowing the host DM to evolve with redshift (\emph{bottom right}). None of these modifications resolve the discrepancy identified in Figure~\ref{fig:catalog_modelprediction_comparison}. As in that figure, the region $\mathcal{R}$ highlighted here contains only $\sim15\%$ of the observed sample.}
    \label{fig:poisson_result}
\end{figure*}

\begin{deluxetable*}{lcccccccc} % 9 columns: Model, Phi0, n, gamma, Delta gamma, Echar, Ebreak, mu_tau, sigma_tau
\tabletypesize{\footnotesize}
\tablewidth{0pt}
\renewcommand{\arraystretch}{1.35}
\setlength{\tabcolsep}{2.2pt}
\tablecaption{Best-fit parameters and $68\%$ credible intervals across the
fiducial $R(F,{\rm DM}|\boldsymbol{\lambda})$ model, compared with results by S23, and the extended models described in Appendix~\ref{app:supplement_results}. 
Values represent the medians of the posterior distributions, with uncertainties derived from the 16th and 84th percentiles.
Part 1 of 2; see Table~\ref{tab:mcmc_master_bottom} for $\alpha$, $\mu_{\rm host,10}$, $\sigma_{\rm host,10}$, $n_z$, $p$-value.
\label{tab:mcmc_master_top}}
\tablehead{
\colhead{Model} & \colhead{$\log_{10}\Phi_0$\,\tablenotemark{a}} & \colhead{$n$} & \colhead{$\gamma$\,\tablenotemark{b}} & \colhead{$\Delta\gamma$}
& \colhead{$\log_{10}E_{\rm char}$\,\tablenotemark{a}} & \colhead{$\log_{10}E_{\rm break}$\,\tablenotemark{a}} & \colhead{$\mu_{\tau,10}$\,\tablenotemark{c}} & \colhead{$\sigma_{\tau,10}$\,\tablenotemark{c}}
}
\startdata
Uniform prior range & $[-0.96,\,6.43]$ & $[-2.00,\,8.00]$ & $[-3.00,\,3.00]$ & $[0.00,\,8.00]$ & $[38.00,\,49.00]$ & $[35.00,\,49.00]$ & $[-5.00,\,1.15]$ & $[0.004,\,1.30]$ \\[0.3em]
\tableline
S23 & $4.9^{+0.3}_{-0.3}$ & $1.0^{+0.8}_{-0.7}$ & $-1.3^{+0.7}_{-0.4}$ & $\ldots$ & $41.4^{+0.5}_{-0.5}$ & $\ldots$ & $\ldots$ & $\ldots$ \\[0.3em]
Base model & $5.06^{+0.15}_{-0.16}$ & $0.5^{+0.4}_{-0.4}$ & $-1.41^{+0.27}_{-0.21}$ & $\ldots$ & $41.3^{+0.4}_{-0.3}$ & $\ldots$ & $\ldots$ & $\ldots$ \\[0.3em]
Rate interpretation & $4.98^{+0.15}_{-0.16}$ & $0.9^{+0.7}_{-0.6}$ & $-1.34^{+0.27}_{-0.21}$ & $\ldots$ & $41.18^{+0.28}_{-0.25}$ & $\ldots$ & $\ldots$ & $\ldots$ \\[0.3em]
Delay-time evolution & $4.99^{+0.15}_{-0.17}$ & $\ldots$ & $-1.58^{+0.15}_{-0.13}$ & $\ldots$ & $41.68^{+0.16}_{-0.16}$ & $\ldots$ & $-2.5^{+1.7}_{-1.7}$ & $0.7^{+0.4}_{-0.4}$ \\[0.3em]
Broken power-law & $5.13^{+0.13}_{-0.16}$ & $0.21^{+0.4}_{-0.4}$ & $-1.58^{+0.16}_{-0.20}$ & $5.0^{+1.9}_{-1.9}$ & $\ldots$ & $41.4^{+0.3}_{-0.2}$ & $\ldots$ & $\ldots$ \\[0.3em]
Evolving DM$_{\rm host}$ & $5.30^{+0.22}_{-0.21}$ & $0.8^{+0.8}_{-0.6}$ & $-1.61^{+0.21}_{-0.17}$ & $\ldots$ & $41.3^{+0.4}_{-0.4}$ & $\ldots$ & $\ldots$ & $\ldots$ \\[0.3em]
\enddata
\tablenotetext{a}{Fit in natural log-space, presented here in base-10.}
\tablenotetext{b}{{Tabulated here are the statistical uncertainties resulting from the MCMC framework. We estimate the systematic uncertainties to be $\pm25\%$ in \S\ref{sec:results}, which are  dominant as compared to the statistical uncertainties for $\gamma$, $\mu_{{\rm host},10}$, and $\sigma_{{\rm host},10}$; hence, we utilize the systematic uncertainty values for the analyses presented in the paper}.}
\tablenotetext{c}{Location and scale parameters of the single-channel lognormal delay-time distribution, i.e., the mean and standard deviation of $\ln(\tau/{\rm Gyr})$ (Appendix~\ref{app:supplement_results}). Fit in natural-log space and presented here in base 10 (e.g., $\mu_{\tau,10}=\mu_\tau\log_{10}e$).}
\end{deluxetable*}

\begin{deluxetable*}{lcccccl} % 7 columns: Model, alpha, mu, sigma, n_z, p-value, Comments (wraps via \parbox in the cells, not the column spec)
\tabletypesize{\footnotesize}
\tablewidth{0pt}
\setlength{\tabcolsep}{6.0pt}
\tablecaption{Continued from Table~\ref{tab:mcmc_master_top}: $\alpha$, $\mu_{\rm host,10}$, $\sigma_{\rm host,10}$, $n_z$,
goodness-of-fit $p$-value, and comments for each model. Goodness-of-fit $p$-value for the model for the region $\mathcal{R}$, and its complement $\mathcal{R}^c$, described in \S\ref{sec:goodness_of_fit}. The $p$-value upper limit is determined by the trial factor of $10^4$ (see \S\ref{sec:goodness_of_fit}). None of the alternate models alleviate the tension with the observed data in \catii.
\label{tab:mcmc_master_bottom}}
\tablehead{
\colhead{Model} & \colhead{$\alpha$} & \colhead{$\mu_{\rm host,10}$\,\tablenotemark{a,b}} & \colhead{$\sigma_{\rm host,10}$\,\tablenotemark{a,b}}
& \colhead{$n_z$} & \colhead{$(\mathcal{R}^c,\mathcal{R})\;p$-value} & \colhead{Comments}
}
\startdata
Uniform prior range & $[-5.00,\,5.00]$ & $[1.30,\,2.70]$ & $[0.04,\,1.74]$ & $[-2.00,\,5.00]$ & --- & --- \\[0.3em]
\tableline
S23 & $-1.4^{+0.9}_{-1.2}$ & $1.93^{+0.26}_{-0.38}$ & $0.41^{+0.21}_{-0.20}$ & $\ldots$ & $\ldots$ & \parbox[t]{4.0cm}{Results from \cite{smb+23} using CHIME/FRB Catalog 1.} \\[0.3em]
Base model & $-1.0^{+0.5}_{-0.6}$ & $2.30^{+0.07}_{-0.09}$ & $0.36^{+0.04}_{-0.04}$ & $\ldots$ & $(0.16,<10^{-4})$ & \parbox[t]{4.0cm}{Model described in \S\ref{sec:pop_model} fit using CHIME/FRB Catalog 2.} \\[0.3em]
Rate interpretation & $-0.7^{+0.4}_{-0.4}$ & $2.30^{+0.07}_{-0.08}$ & $0.36^{+0.04}_{-0.04}$ & $\ldots$ & $(0.08,<10^{-4})$ & \parbox[t]{4.0cm}{Assumes a frequency-dependent rate $\Phi\sim\Phi(z, \nu)= \Phi(z)\nu^{\alpha'}$.} \\[0.3em]
Delay-time evolution & $-1.7^{+0.2}_{-0.2}$ & $2.22^{+0.07}_{-0.08}$ & $0.33^{+0.04}_{-0.04}$ & $\ldots$ & {$(0.05,<10^{-4})$} & \parbox[t]{4.0cm}{Single channel with a lognormal delay-time distribution. $\mu_{\tau,10}$ and $\sigma_{\tau,10}$ remain unconstrained.} \\[0.3em]
Broken power-law & $-0.7^{+0.5}_{-0.6}$ & $2.34^{+0.06}_{-0.08}$ & $0.35^{+0.03}_{-0.04}$ & $\ldots$ & $(0.11,<10^{-4})$ & \parbox[t]{4.0cm}{Energy function modeled as a broken power-law. $\Delta\gamma$ remains unconstrained.} \\[0.3em]
Evolving DM$_{\rm host}$ & $-1.7^{+0.9}_{-1.6}$ & $2.32^{+0.07}_{-0.08}$ & $0.29^{+0.06}_{-0.04}$ & $1.6^{+1.4}_{-1.6}$ & $(0.06,<10^{-4})$ & \parbox[t]{4.0cm}{Allows for a redshift evolution of DM host with redshift following \cite{llu26}.} \\[0.3em]
\enddata
\end{deluxetable*}

\end{document}